\documentclass[twocolumn,aps,prd,nofootinbib,superscriptaddress,showpacs]{revtex4-1}
\usepackage{amsmath,graphicx,bm,amsbsy,aas_macros,color,url,natbib}

\newcommand{\BigO}[1]{\mathcal{O}(#1)}
\newcommand{\F}{\mathbf{F}}
\newcommand{\M}{\mathbf{M}}
\newcommand{\C}{\mathbf{C}}
\newcommand{\beq}{\begin{equation}}
\newcommand{\eeq}{\end{equation}}

\newcommand{\Eye}{\mathbf{I}}
\newcommand{\x}{\mathbf{x}}

\begin{document} 

\title{Overcoming real-world obstacles in 21\,cm power spectrum estimation: A method demonstration and results from early Murchison Widefield Array data}

\author{Joshua S. Dillon}
\email{Because the contributions of the first two authors were essentially equal, we determined their order by a Monte Carlo algorithm inspired by the one presented in Section \ref{sec:DataVolume}.  They can be reached for questions or comments at \url{jsdillon@mit.edu} and \url{acliu@berkeley.edu}.}
\affiliation{MIT Kavli Institute, Massachusetts Institute of Technology, Cambridge, MA, USA}
\affiliation{Dept. of Physics, Massachusetts Institute of Technology, Cambridge, MA, USA}

\author{Adrian Liu}
\email{Because the contributions of the first two authors were essentially equal, we determined their order by a Monte Carlo algorithm inspired by the one presented in Section \ref{sec:DataVolume}.  They can be reached for questions or comments at \url{jsdillon@mit.edu} and \url{acliu@berkeley.edu}.}
\affiliation{MIT Kavli Institute, Massachusetts Institute of Technology, Cambridge, MA, USA}
\affiliation{Dept. of Physics, Massachusetts Institute of Technology, Cambridge, MA, USA}
\affiliation{Dept. of Astronomy and Berkeley Center for Cosmological Physics, Berkeley, CA, USA}

\author{Christopher L. Williams}
\affiliation{MIT Kavli Institute, Massachusetts Institute of Technology, Cambridge, MA, USA}
\affiliation{Dept. of Physics, Massachusetts Institute of Technology, Cambridge, MA, USA}

\author{Jacqueline N. Hewitt}
\affiliation{MIT Kavli Institute, Massachusetts Institute of Technology, Cambridge, MA, USA}
\affiliation{Dept. of Physics, Massachusetts Institute of Technology, Cambridge, MA, USA}

\author{Max Tegmark}
\affiliation{MIT Kavli Institute, Massachusetts Institute of Technology, Cambridge, MA, USA}
\affiliation{Dept. of Physics, Massachusetts Institute of Technology, Cambridge, MA, USA}

\author{Edward H. Morgan}
\affiliation{MIT Kavli Institute, Massachusetts Institute of Technology, Cambridge, MA, USA}

\author{Alan M. Levine}
\affiliation{MIT Kavli Institute, Massachusetts Institute of Technology, Cambridge, MA, USA}

\author{Miguel F. Morales}
\affiliation{Dept. of Physics, University of Washington, Seattle, WA, USA}

\author{Steven J. Tingay}
\affiliation{International Centre for Radio Astronomy Research, Curtin University, Perth, WA, Australia}
\affiliation{ARC Centre of Excellence for All-sky Astrophysics (CAASTRO), Australia}

\author{Gianni Bernardi}
\affiliation{Harvard-Smithsonian Center for Astrophysics, Harvard University, Cambridge, MA, USA}

\author{Judd D. Bowman}
\affiliation{School of Space and Earth Exploration, Arizona State University, Tempe, AZ, USA}

\author{Frank H. Briggs}
\affiliation{ARC Centre of Excellence for All-sky Astrophysics (CAASTRO), Australia}
\affiliation{Research School of Astronomy and Astrophysics, The Australian National University, Canberra, Australia}

\author{Roger C. Cappallo}
\affiliation{MIT-Haystack Observatory, Westford, MA, USA}

\author{David Emrich}
\affiliation{International Centre for Radio Astronomy Research, Curtin University, Perth, WA, Australia}

\author{Daniel A. Mitchell}
\affiliation{ARC Centre of Excellence for All-sky Astrophysics (CAASTRO), Australia}
\affiliation{School of Physics, The University of Melbourne, Melbourne, Australia}

\author{Divya Oberoi}
\affiliation{MIT-Haystack Observatory, Westford, MA, USA}
\affiliation{National Centre for Radio Astrophysics, Tata Institute of Fundamental Research, Pune, India}

\author{Thiagaraj Prabu}
\affiliation{Raman Research Institute, Bangalore, India}

\author{Randall Wayth}
\affiliation{International Centre for Radio Astronomy Research, Curtin University, Perth, WA, Australia}
\affiliation{ARC Centre of Excellence for All-sky Astrophysics (CAASTRO), Australia}

\author{Rachel L. Webster}
\affiliation{ARC Centre of Excellence for All-sky Astrophysics (CAASTRO), Australia}
\affiliation{School of Physics, The University of Melbourne, Melbourne, Australia}

\date{January 15, 2014}

\pacs{95.75.-z, 95.85.Bh, 98.62.Ra, 98.80.-k, 98.80.Es}

\begin{abstract}
We present techniques for bridging the gap between idealized inverse covariance weighted quadratic estimation of 21 cm power spectra and the real-world challenges presented universally by interferometric observation.  By carefully evaluating various estimators and adapting our techniques for large but incomplete data sets, we develop a robust power spectrum estimation framework that preserves the so-called ``EoR window" and keeps track of estimator errors and covariances.  We apply our method to observations from the 32-tile prototype of the Murchinson Widefield Array to demonstrate the importance of a judicious analysis technique.  Lastly, we apply our method to investigate the dependence of the clean EoR window on frequency---especially the frequency dependence of the so-called ``wedge" feature---and establish upper limits on the power spectrum from $z=6.2$ to $z=11.7$.  Our lowest limit is $\Delta(k) < 0.3$ Kelvin at 95\% confidence at a comoving scale $k = 0.046$ Mpc$^{-1}$ and $z = 9.5$.
\end{abstract}

\maketitle

\section{Introduction}
\label{sec:Intro}

In recent years, $21\,\textrm{cm}$ tomography has emerged as a promising probe of the Epoch of Reionization (EoR).  As a direct measurement of the three-dimensional distribution of neutral hydrogen at high redshift, the technique will allow detailed study of the complex astrophysical interplay between the intergalactic medium and the first luminous structures of our Universe.  This will eventually pave the way towards the use of $21\,\textrm{cm}$ tomography to constrain cosmological parameters to exquisite precision, thanks to the enormity of the physical space within its reach (please see, e.g., \citet{FurlanettoReview,miguelreview,PritchardLoebReview,aviBook} for recent reviews).

To date, observational efforts have focused on measurements of the $21\,\textrm{cm}$ power spectrum.  Such a measurement is exceedingly difficult.  Sensitivity requirements are extreme, requiring thousands of hours of integration and large collecting areas \citep{MiguelNoise,Judd06,LidzRiseFall,LOFAR2,AaronSensitivity}.  Adding to this challenge is the fact that raw sensitivity is insufficient---what counts is sensitivity to the cosmological signal above expected contaminants like galactic synchrotron radiation, which are three to four orders of magnitude brighter at the relevant frequencies \citep{Angelica,LOFAR,BernardiForegrounds,PoberWedge}. 
 
\begin{figure}[!ht] 
	\centering 
	\includegraphics[width=.49\textwidth]{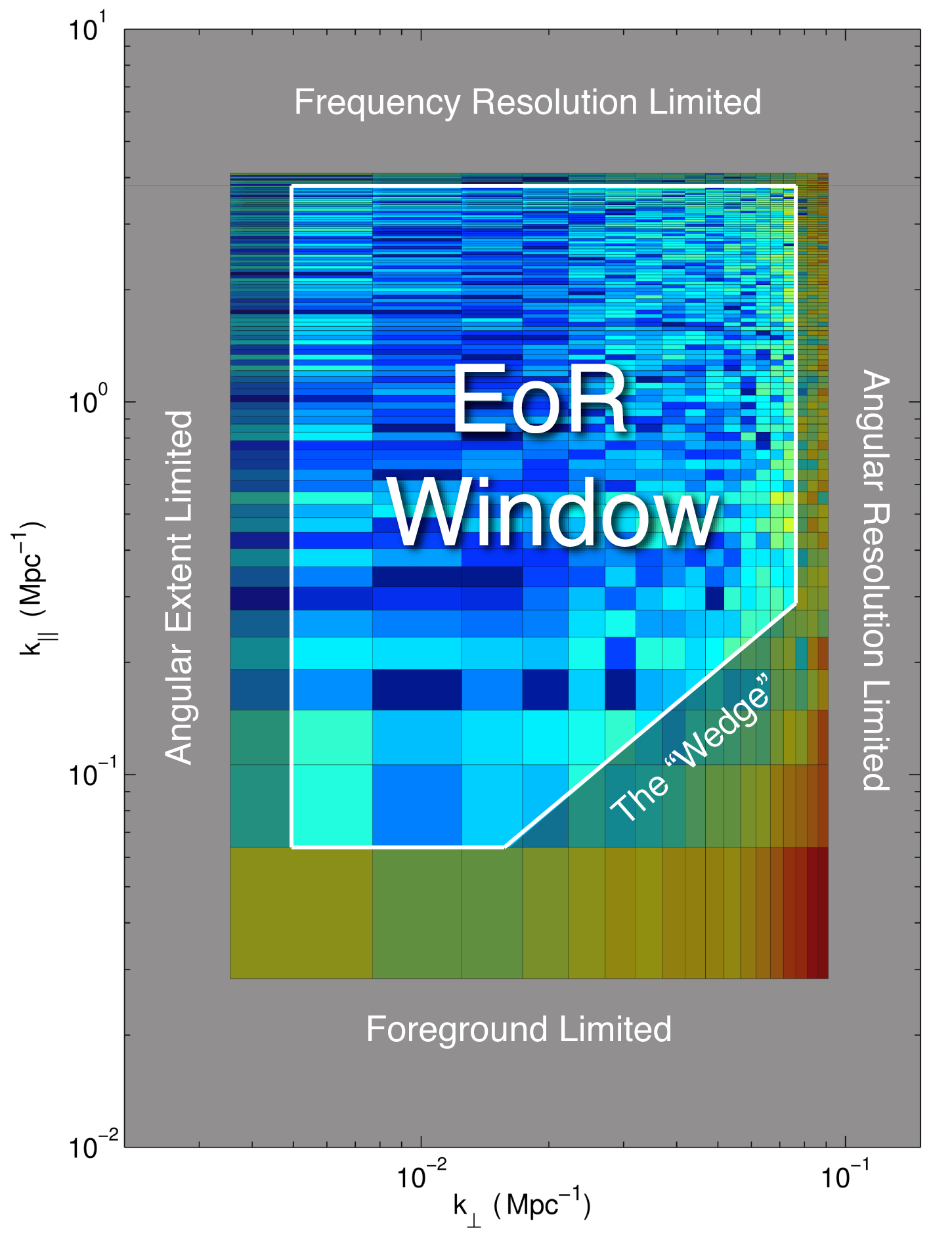}
	\caption{The ``EoR window," a region of Fourier space with relatively low noise and foregrounds, is thought to present the best opportunity for measuring the cosmological $21\,\textrm{cm}$ power spectrum during the Epoch of Reionization.  Here we show an example power spectrum from early MWA data, as a function of $k_\perp$ (Fourier components perpendicular to the line of sight) and $k_\parallel$ (Fourier components parallel to the line of sight).  More details on how we have calculated and plotted $P(k_\perp,k_\parallel)$ are found in Section \ref{sec:WorkedExample}.  We schematically highlight the instrumental and foreground effects that that delimit the EoR window---the coldest part of this power spectrum.  At low and high $k_\perp$, measurements are limited by an instrument's ability to probe the largest and smallest angular scales, respectively.  Limited spectral resolution causes similar effects at the highest $k_\parallel$.  As spectrally smooth sources, foregrounds inhabit primarily the low $k_\parallel$ regions.  Thanks to chromatic instrumental effects, however, there is a slight encroachment of foregrounds towards higher $k_\parallel$ at higher $k_\perp$, in what has been colloquially termed the ``wedge" feature.}
	\label{fig:EoRWindow}
\end{figure}

To deal with these challenges, numerous techniques have been proposed and implemented for foreground mitigation and power spectrum estimation.  These include foreground removal via parametric fits \citep{xiaomin,Judd08,paper2,paper1}, non-parametric methods \citep{Harker,Chapman1,Chapman2}, principal component analyses \citep{GMRT,AdrianForegrounds,GBT,newGMRT}, filtering \citep{nusserforegrounds,PetrovicOh,AaronDelay}, frequency stacking \citep{ChoForegrounds}, and quadratic methods \citep{LT11,DillonFast,Richard}.  In almost all of these proposals, foregrounds are separated from the cosmological signal by taking advantage of the differences in their spectra.  Foregrounds are dominated by continuum processes and thus have smooth spectra.  On the other hand, because the cosmological line-of-sight distance maps to the observed frequency of the redshifted $21\,\textrm{cm}$ line, the rapid fluctuations in the brightness temperature distribution that are expected from theory will map to a measured cosmological signal with jagged, rapidly fluctuating spectra.  When these spectral differences are considered in conjunction with instrumental characteristics, one can identify an ``EoR window": a region in Fourier space where power spectrum measurements are expected to be relatively free from foregrounds \citep{Dattapowerspec,AaronDelay,VedanthamWedge,MoralesPSShapes,Trottwedge,NithyaSubmitted}.  This is shown schematically in Figure \ref{fig:EoRWindow}, where we have used early Murchison Widefield Array (MWA) data to estimate the power spectrum as a function of $k_\perp$ (Fourier mode perpendicular to the line-of-sight) and $k_\parallel$ (Fourier mode parallel to the line-of-sight).  
More details regarding this figuew are provided in Section \ref{sec:WorkedExample}; for now we simply wish to draw attention to the existence of a relatively contaminant-free region in the middle of the $k_\perp$-$k_\parallel$ plane.  This clean region is what we denote the EoR window.

The EoR window is generally considered the sweet spot for an initial detection of the cosmological $21\,\textrm{cm}$ power spectrum, and constraints are likely to degrade away from the window.  At high $k_\perp$ (i.e., the finest angular features on the sky), errors increase due to the angular resolution limitations of one's instrument.  For an interferometer, this resolution is roughly set by the length of the longest baseline.  Conversely, the shortest baselines define the largest modes that are observable by the instrument.  Errors therefore also increase at the lowest $k_\perp$ where again there are few baselines.

A similar limitation defines the boundary of the EoR window at high $k_\parallel$.  Since the spectral nature of $21\,\textrm{cm}$ measurements mean that different observed frequencies map to different redshifts, the highest $k_\parallel$ modes are inaccessible due to the limited spectral resolution of one's instrument.  At low $k_\parallel$, one probes spectrally smooth modes---precisely those that are expected to be foreground contaminated.  Thus there is another boundary to the EoR window at low $k_\parallel$.

A final delineation of the EoR window is provided by the region labeled as the ``wedge" in Figure \ref{fig:EoRWindow}.  The wedge feature is a result of an interplay between angular and spectral effects.  Simulations have shown that the wedge is the effect of chromaticity in one's synthesized beam (which is inevitable when an interferometer is used to survey the sky).  This chromaticity imprints unsmooth spectral features on measured foregrounds, resulting in foreground contamination beyond the lowest $k_\parallel$ modes even if the foregrounds themselves are spectrally smooth.  Luckily, this sort of additional contamination follows a reasonably predictable pattern in the $k_\perp$-$k_\parallel$ plane, and in the limit of intrinsically smooth foregrounds, the wedge can be shown to extend no farther than the line
\begin{equation}
\label{eq:wedge}
k_\parallel = \left[ \sin \theta_{\textrm{field}} \frac{D_M (z) E(z)}{D_H (1+z)} \right] k_\perp,
\end{equation}
where $D_H \equiv c/H_0$, $E(z) \equiv \sqrt{\Omega_m (1+z)^3 + \Omega_\Lambda}$, $D_M(z) \equiv \int_0^z dz^\prime / E(z^\prime)$, $\theta_{\textrm{field}}$ is angular radius of the the field-of-view, and $c$, $H_0$, $\Omega_m$, and $\Omega_\Lambda$ have their usual meanings \citep{Dattapowerspec,VedanthamWedge,MoralesPSShapes,Trottwedge}. Intuitively, the foreground-contaminated wedge extends to higher $k_\parallel$ at higher $k_\perp$ because the high $k_\perp$ modes are probed by the longer baselines of an interferometer array, which have higher fringe rates that more effectively imprint spectral structure in the measured signals.  For an alternate but equivalent explanation in terms of delay modes, please see the illuminating discussion in \citet{AaronDelay}.

The concept of an EoR window is important in that it provides relatively strict boundaries that separate fairly foreground-free regions of Fourier space from heavily foreground-contaminated ones.  It therefore provides one with the option of practicing foreground avoidance rather than foreground subtraction.  If it turns out that foregrounds cannot be modeled well enough to be directly subtracted with the level of precision required to detect the cosmological signal, foreground avoidance becomes an important alternative, in that the only way to robustly suppress foregrounds is to preferentially make measurements within the EoR window.  Likely, some combination of the two strategies---foreground subtraction and foreground avoidance---will prove useful for the detection of the 21 cm power spectrum. Of course, measurements within the EoR window are still contaminated by instrumental noise, but fortunately the noise integrates down with further observation time (as long as calibration errors and other instrumental systematics can be sufficiently minimized).  Observationally, it is encouraging that the EoR window has now been shown to be free of foregrounds to better than one part in a hundred in power \citep{PoberWedge}.  

As experimental sensitivities increase, however, one must take care to preserve the cleanliness of the EoR window to an even higher dynamic range.  There are several ways in which our notion of the EoR window may be compromised.  First, as experiments integrate in time and acquire greater sensitivity, we may discover that our approximation of spectrally smooth foregrounds is insufficiently good for a detection of the (faint) cosmological signal.  In other words, foreground sources may have small but non-negligible high $k_\parallel$ components in their spectra that have thus far gone undetected.  This would translate into a smaller-than-expected EoR window.  In addition, even intrinsically smooth foregrounds may appear jagged in a real measurement because of instrumental effects such as imperfect calibration.  The precise interferometer layout may also result in unsmooth artifacts that arise from combining data from non-redundant baselines \citep{Hazelton2013}.  Finally, suppose that the aforementioned effects are negligible and that the assumption of spectrally smooth foreground emission continues to hold.  The EoR window still cannot be taken for granted because non-optimal data analysis techniques may result in unwanted foreground artifacts in the region.   For the EoR window to exist at all, it is essential that power spectra are estimated in a rigorous fashion, with well-understood statistics.  

The goal of this paper is to minimize unwanted data analysis artifacts by establishing methods for power spectrum estimation that are both robust and as optimal as possible.  Previous efforts have rarely met both criteria: either the methods are robustly applicable to data with real-world artifacts but fail to achieve optimized (or even rigorously computable) error properties, or provide an optimal framework but ignore real-world complications.  In this paper we extend the rigorous framework described in \citet{LT11} and \citet{DillonFast} to deal with real-world effects.  The result is a computationally feasible approach to analyzing real data that not only preserves the cleanliness of the EoR window, but also rigorously keeps track of all relevant error statistics.

To demonstrate the applicability of our approach, we apply our techniques to early data from the Murchison Widefield Array (MWA).  These data were derived from $\sim 22\,\textrm{hours}$ of tracked observations using an early, 32-element prototype array.  The results are therefore not designed to be cosmologically competitive, but instead illustrate the rigor that will be required for an eventual detection of the EoR while also providing new measurements on the ``wedge" feature that delineates the EoR window.

This paper is organized as follows.  In Section \ref{sec:Methods} we discuss various real-world obstacles that must be dealt with when analyzing real data, and how one can overcome them while maintaining statistical rigor.  We then apply our methods to MWA data in Section \ref{sec:WorkedExample} as a ``worked example", highlighting the importance of various subtleties of power spectrum estimation.  In Section \ref{sec:earlyResults} we present some results from the data, emphasizing the agreement between theoretical expectations and our observations of the foreground wedge (particularly regarding the frequency dependence of the wedge).  We also present upper limits on the cosmological $21\,\textrm{cm}$ power spectrum over the broad redshift range of $z=6.2$ to $z=11.1$.  Finally, we summarize our conclusions in Section \ref{sec:Conc}.
 
\section{Systematic methods for dealing with real-world obstacles}
\label{sec:Methods}

To understand the gap between an analysis framework for idealized observations and any real-world data set, we enumerate and address six different obstacles that rather universally affect real data.  Our goal in this section is to meet the challenges presented by these obstacles while maintaining as many of the advantages of the optimal framework as possible, which we reiterate in Section \ref{sec:IdealObs}, especially the ability to minimize and precisely quantify the uncertainties in the measurements.  In the following sections, we address the problems presented by large data volumes (Section \ref{sec:DataVolume}), uncertainties in the properties of contaminants such as foregrounds (Section \ref{sec:crossPower}), incomplete $uv$ coverage (Section \ref{sec:IncompleteUV}), radio frequency interference (RFI) flagging (Section \ref{sec:RFI}), foreground leakage into the EoR window (Section \ref{sec:decorr}), and binning to spherically averaged power spectra (Section \ref{sec:cylindToSph}).

\subsection{A systematic framework for analyzing idealized observations}
\label{sec:IdealObs}
In this section, we briefly review the formalism of \citet{LT11} for optimal power spectrum estimation, which was adapted for $21\,\textrm{cm}$ tomography from similar techniques used in galaxy survey and cosmic microwave background analysis \citep{Maxpowerspeclossless,BJK,Maxgalaxysurvey1,Maxgalaxysurvey2}.  For now, we do not include real-world effects such as missing data from RFI flagging, and the purpose of later sections is to extend the formalism to take into account these complications.

In $21\,\textrm{cm}$ tomography, one typically wishes to measure both the spherically-binned power spectrum $P_{\textrm{sph}}(k)$, defined by
\begin{equation}
\langle \widetilde{T}^* (\mathbf{k}) \widetilde{T} (\mathbf{k}^\prime) \rangle \equiv (2 \pi)^3 P_{\textrm{sph}}(k) \delta (\mathbf{k} - \mathbf{k}^\prime),
\end{equation}
and the cylindrically-binned power spectrum $P_{\textrm{cyl}} (k_\perp, k_\parallel)$, defined by
\begin{equation}
\langle \widetilde{T}^* (\mathbf{k}) \widetilde{T} (\mathbf{k}^\prime) \rangle \equiv (2 \pi)^3 P_{\textrm{cyl}}(k_\perp, k_\parallel) \delta (\mathbf{k} - \mathbf{k}^\prime),
\end{equation}
with $\widetilde{T} (\mathbf{k})$ signifying the spatial Fourier transform of the $21\,\textrm{cm}$ brightness temperature field $T(\mathbf{r})$, $\mathbf{k}$ denoting the spatial wavevector with magnitude $k$, and components $k_\perp$ and $k_\parallel$ as the components perpendicular and parallel to the line-of-sight, respectively.  The angled brackets $\langle \cdots \rangle$ represent an ensemble average.  The spherical power spectrum is useful for comparing to theoretical models, since it is obtained by angularly averaging over spherical shells in Fourier space, and thus makes the cosmologically relevant assumption of isotropy.  The cylindrical power spectrum is useful for identifying instrumental and foreground effects, which possess a cylindrical symmetry rather than a spherical one.  Typically, the cylindrical power spectrum is produced first as a tool for foreground isolation (i.e., to identify the EoR window), and then subsequently binned into a spherical power spectrum.  This section concerns the estimation of the cylindrical power spectrum.  Optimal binning techniques to go from the cylindrical spectrum to the spherical spectrum are discussed in Section \ref{sec:cylindToSph}.

In estimating a power spectrum from data, one must necessarily discretize the problem.  We make the approximation that the power spectra are piecewise constant functions, such that we can describe them in terms of a vector of bandpowers with components $p^\alpha$, where
\begin{equation}
p_\alpha \equiv P_{\textrm{cyl}} (k^\alpha_\perp, k^\alpha_\parallel).
\end{equation}
It is the bandpowers and their error properties that one wishes to estimate from the data, which come in the form of a data vector $\mathbf{x}$.  Intuitively, one can think of the data vector as a list of the $21\,\textrm{cm}$ brightness temperatures measured at various locations in a three-dimensional ``data cube".  Rigorously, we define each element of the data vector (i.e., each voxel of the data cube) as
\begin{equation}
\label{eq:pixDef}
\mathbf{x}_i \equiv \int T(\mathbf{r}) \psi_i (\mathbf{r}) d^3 \mathbf{r},
\end{equation}
with $\psi_i (\mathbf{r})$ being the pixelization kernel and $T(\mathbf{r})$ as the (continuous) three-dimensional $21\,\textrm{cm}$ brightness temperature field\footnote{Of course, instrumental noise and foregrounds do not properly reside in a cosmological three-dimensional volume: noise is introduced in the electronics of the system, whereas foregrounds are ``nearby" and only appear in the same location in the data cube as our cosmological signal by virtue of their frequency dependence.  However, there is a gain in convenience and no loss of generality in assigning a noise and foreground contribution to each voxel, pretending that those contaminants also live in the observed cosmological volume.}.  In this paper we take the $i^{th}$ pixelization kernel $\psi_i (\mathbf{r})$ to be a boxcar function centered on the $i^{th}$ voxel of the data.\footnote{This choice, following \cite{DillonFast}, is motivated by the fact that the covariance between each pixel in this basis for both noise and foregrounds can be written in an algorithmically convenient way.}

To estimate the $\alpha^{th}$ bandpower from the data vector, we first form a quadratic estimator of the form
\begin{align}
\label{eq:unnormBands}
q_\alpha \equiv& \frac{1}{2}( \mathbf{x} - \mathbf{m} )^t \mathbf{C}^{-1} \mathbf{C}_{,\alpha} \mathbf{C}^{-1}( \mathbf{x} -\mathbf{m} ) \nonumber \\ 
&- \frac{1}{2} \textrm{tr} [ \mathbf{C}_{\textrm{junk}} \mathbf{C}^{-1} \mathbf{C}_{,\alpha} \mathbf{C}^{-1} ],
\end{align}
where $\mathbf{m} \equiv \langle \mathbf{x} \rangle$ is the mean of the data, $\mathbf{C} \equiv \langle \mathbf{x} \mathbf{x}^t \rangle - \langle \mathbf{x} \rangle \langle \mathbf{x} \rangle^t$ is its covariance, $\mathbf{C}_{\textrm{junk}}$ is the component of the covariance ``junk"/contaminants (to be defined in the following section), and $\mathbf{C}_{,\alpha}$ is the derivative of the covariance with respect to the $\alpha^{th}$ bandpower.  Since we are approximating the power spectrum as piecewise constant, we have
\begin{equation}
\mathbf{C} =\mathbf{C}_{\textrm{junk}} +  \sum_{\alpha} p_\alpha \mathbf{C}_{,\alpha}.
\end{equation}
Combined with Equation \eqref{eq:pixDef}, this expression can be used to derive explicit forms for $\mathbf{C}_{,\alpha}$, which reveals that the matrix essentially Fourier transforms and bins the data \citep{LT11,DillonFast}.  Intuitively, $\mathbf{C}_{,\alpha}$ can be thought of as the response in the data covariance $\mathbf{C}$ to the bandpower $p_\alpha$.  Thus, as long as one selects  an appropriate form for $\mathbf{C}_{,\alpha}$, the formalism of this section can also be used to directly measure the spherical power spectrum.  However, as we discussed above, in this paper we choose to first estimate the cylindrical power spectrum as an intermediate diagnostic step, to quantify and mitigate foregrounds better.
 
Once the $q_\alpha$s have been formed, they need to be normalized using a suitable invertible matrix $\mathbf{M}$ to form the final bandpower estimates:
\begin{equation}
\label{eq:normingBands}
\mathbf{\widehat{p}} = \mathbf{M} \mathbf{q},
\end{equation}
where we have grouped the bandpower estimates $\widehat{p}_\alpha$ into a vector $\widehat{\mathbf{p}}$ (and similarly grouped the coefficients $q_\alpha$ and $\mathbf{q}$), 
with the hat ( $\widehat{ }$ ) signifying the fact that we have formed an \emph{estimator} of the true bandpowers\footnote{Note that $\mathbf{q}$, $\mathbf{\widehat{p}}$, and $\mathbf{M}$ live in a different vector space than $\mathbf{x}$, $\mathbf{C}$, and $\mathbf{C}_{,\alpha}$.  The former are in a vector space where each component refers to a different bandpower, whereas the latter are in one where different components refer to different voxels.}.  We shall discuss different choices of $\mathbf{M}$ in Section \ref{sec:decorr}.

To understand the uncertainty in our estimates, we compute several error properties.  The first is the covariance matrix of the final measured bandpowers:
\begin{equation}
\label{eq:CovP}
\boldsymbol \Sigma \equiv \langle \mathbf{\widehat{p}} \mathbf{\widehat{p}}^t \rangle - \langle \mathbf{\widehat{p}} \rangle \langle \mathbf{\widehat{p}} \rangle^t = \mathbf{M} \mathbf{F} \mathbf{M}^t,
\end{equation}
where we have introduced the Fisher matrix $\mathbf{F}$, which has components
\begin{equation}
F_{\alpha \beta} = \frac{1}{2} \textrm{tr}[ \mathbf{C}^{-1} \mathbf{C}_{, \alpha} \mathbf{C}^{-1} \mathbf{C}_{,\beta} ]. \label{eq:fisherDef}
\end{equation}
The Fisher matrix also allows us to relate our estimated bandpowers $\mathbf{\widehat{p}}$ to the true bandpowers $\mathbf{p}$ via the window function matrix $\mathbf{W}$:
\begin{equation}
\label{eq:phatWp}
\langle \mathbf{\widehat{p}} \rangle = \mathbf{W} \mathbf{p},
\end{equation}
where $\mathbf{W}$ can be shown to take the form
\begin{equation}
\mathbf{W} = \mathbf{M} \mathbf{F}.
\end{equation}
If we choose $\mathbf{M}$ such that the rows of $\mathbf{W}$ each sum to unity, Equation \eqref{eq:phatWp} shows that each bandpower estimate can be thought of as a weighted average of the truth, with weights given by each row (each window function).  Even with this normalization requirement, there are still many choices for $\mathbf{M}$.  We discuss the various options and tradeoffs in Section \ref{sec:decorr}.

Whatever the choice of $\mathbf{M}$, our estimator has optimal error properties in the sense that if $\mathbf{\widehat{p}}$ in Equation \eqref{eq:phatWp} is used to constrain parameters in some theoretical model, those measured parameters will have the smallest possible error bars given the observed data \citep{Maxpowerspeclossless}.  Our goal in the following sections will be to ensure that both these small error bars and our ability to rigorously compute them are preserved in the face of real-world difficulties.

\subsection{A real-world obstacle: data volume}
\label{sec:DataVolume} 
Perhaps the most glaring difficulty presented by the ideal technique outlined above is its computational cost.  Much of that cost arises from the inversion of the data covariance matrix $\mathbf{C}$ in Equations \eqref{eq:unnormBands} and \eqref{eq:fisherDef}, in addition to the multiplication of $\C$ and matrices of the same size.  Both of these operations scale like $\BigO{N^3}$, where $N$ is the number of voxels in each data vector.  The computational cost makes taking full advantage of current generational interferometric data prohibitive, not to mention upcoming observational efforts that expect to produce $10^6$ or more voxels of data. 

One would like to retain the information theoretic advantages of the quadratic estimator method and its ability to precisely model errors and window functions, without $\BigO{N^3}$ complexity.  The solution to this problem, developed and demonstrated in \cite{DillonFast}, comes from taking advantage of a number of symmetries and approximate symmetries of the survey geometry and the covariance matrix, $\C$, and can accelerate the technique to $\BigO{N\log N}$.

The fast method relies on assembling the data into a data cube with rectilinear voxels amenable to manipulation with the Fast Fourier Transform.  This is equivalent to the assertion that each voxel represents an equal volume of comoving space, an approximation that relies on two restrictions on the data cube geometry.  First, the range of frequencies considered must be small enough that $D_c(z)$ (the line-of-sight comoving distance, equal to $D_M(z)$ above in a spatially flat universe) is linear with $\nu$.  Generally, one should limit oneself to analyzing the power spectrum of redshift ranges short enough that the evolution of the power spectrum during reionization can be neglected.  This range, suggested by \cite{Yi} to be $\Delta z \lesssim 0.5$, makes the approximation of a linear relationship between $\nu$ and $D_c(z)$ better than one part in $10^3$ at the redshifts of interest to 21 cm cosmology. 

Second, the assumption of equal volume voxels relies on the flat sky approximation.  To achieve this the area surveyed can be broken into a number of subfields, each a few degrees on a side, for which the curvature of the sky can be neglected.  As long as the angular extent of the data cube is smaller than $\sim 10^\circ$, the flat sky approximation is correct to a few parts in $10^3$.

By analyzing a rectilinear volume of the universe, all steps in calculating the band powers $q_\alpha$ can be performed quickly by exploiting various symmetries and taking advantage of the Fast Fourier Transform.  The model for $\C$ can be broken up into a number of independent matrices representing signal, noise, and foregrounds.  Each of these models, developed by \cite{LT11}, is well approximated by a sparse matrix in a convenient combination of real and Fourier spaces \citep{DillonFast}.  As a result, multiplication of a vector by $\C$ can be performed in $\BigO{N\log N}$.  \citet{DillonFast} showed how that speed-up can be parlayed into a method for quickly calculating $q_\alpha$ using the Conjugate Gradient Method.  The rapid convergence of the iterative method for calculating $\C^{-1}\x$ can be ensured by the application of a preconditioner which relies on the spectral smoothness of foregrounds and the fact that they are well described by only a few eigenmodes \citep{AdrianForegrounds}. Then, by randomly simulating many data vectors from the covariance $\C$ and calculating $q_\alpha$ from each, the Fisher matrix can be estimated from the fact that 
\beq
\F = \langle \mathbf{q} \mathbf{q}^t \rangle - \langle \mathbf{q} \rangle \langle \mathbf{q}^t \rangle,
\eeq 
which follows from Equation \eqref{eq:CovP}.  All of this together allows for fast, optimal power spectrum estimation---including error bars and window functions---despite the challenge presented by an enormous volume of data.

\subsection{A real-world obstacle: uncertain contaminant properties}
\label{sec:crossPower}
If one had perfect knowledge of the foreground contamination in the data cube, the problem of foreground contamination would be trivial; one would simply perform a direct subtraction of the foregrounds from the data vector $\mathbf{x}$.  Unfortunately, our knowledge of foregrounds is far from perfect, particularly at the level of precision required for a direct detection of the cosmological $21\,\textrm{cm}$ signal.  Because of this, the estimator shown in Equation \eqref{eq:unnormBands} in fact combines several different foreground subtraction steps in an attempt to achieve the lowest possible level of foreground contamination:
\begin{enumerate}
\item A direct subtraction of a foreground model from the data vector.  This is given by $\mathbf{x} - \mathbf{m}$.  To see this, note that the data vector can be thought of as being comprised of the cosmological $21\,\textrm{cm}$ signal $\mathbf{x}_{21}$, the foregrounds $\mathbf{x}_{\textrm{fg}}$, and the instrumental noise $\mathbf{n}$.  On the other hand, the mean data vector
\begin{equation}
\mathbf{m} \equiv \langle \mathbf{x} \rangle = \langle \mathbf{x}_{21} \rangle + \langle \mathbf{x}_{\textrm{fg}} \rangle + \langle \mathbf{n} \rangle = \langle \mathbf{x}_{\textrm{fg}} \rangle.
\end{equation}
contains only the foreground contribution, because we are interested in the \emph{fluctuations} of the $21\,\textrm{cm}$ signal, so the cosmological signal has zero mean, as does the instrumental noise (in the absence of instrumental systematics).  Note that because the mean here is the mean in the ensemble average sense (as opposed to just the spatial mean), $\mathbf{m}$ represents a full spatial and spectral model of the foregrounds.
\item Since the foregrounds also appear in the covariance matrix, the action of $\mathbf{C}^{-1}$  is to downweight foreground-contaminated modes, exploiting foreground properties such as smooth frequency dependence.
\item Subtracting the term $ \frac{1}{2} \textrm{tr} [ \mathbf{C}_{\textrm{junk}} \mathbf{C}^{-1} \mathbf{C}_{,\alpha} \mathbf{C}^{-1} ]$ eliminates the bias from contaminants.
\item Finally, the binning of the cylindrical power spectrum to the spherical power spectrum provides yet more foreground suppression.  Foregrounds are distributed in select regions on the $k_\perp$-$k_\parallel$ plane (i.e., outside the EoR window) in patterns that do not lie along contours of constant $k = \sqrt{k_\perp^2 + k_\parallel^2}$.  Thus, when binning along such contours to produce a spherical power spectrum, one can selectively downweight parts of the contour with greater foreground contamination, which constitutes a form of foreground cleaning.  Roughly speaking, this corresponds to taking advantage of the fact that foregrounds have a cylindrical symmetry in Fourier space, whereas the signal is spherically isotropic \citep{MiguelJackie1}.  We do note, however, that the formalism we introduce in Section \ref{sec:cylindToSph} is general enough to use any geometric differences between foregrounds and signal.
\end{enumerate}
Of these foreground mitigation strategies, the first and third are direct subtractions (in amplitude and power, respectively), whereas the second and the fourth act through weightings. The former group represent operations that are particularly vulnerable to incorrectly modeled foregrounds.  To see this, recall that the foregrounds are expected to be larger than the cosmological signal by three or four orders of magnitude \citep{Angelica,LOFAR,BernardiForegrounds,PoberWedge}.  Thus, when performing direct subtractions, low-level, unaccounted-for inaccuracies in the foreground model can translate into extremely large biases in the final results.  In addition, significant numerical errors may arise from the subtraction of two large numbers (the data and the foregrounds) to obtain a small number (the measured cosmological signal).

Our goal for the rest of the section is to immunize ourselves against biases from direct subtractions.  Of the direct subtraction steps list above, the Step 1 is likely to be relatively harmless for two reasons.  First, it is immediately followed by the $\mathbf{C}^{-1}$ downweighting.  The downweighting mitigates the effects of inaccuracies in modeling, for the $\mathbf{C}^{-1}$ tends to gives less weight to precisely the modes that have the largest foreground amplitudes, and therefore would be the most susceptible to modeling errors in the first place.  In addition, the uncertainty in foreground properties in those regions of the $k_\perp$-$k_\parallel$ plane result in large error bars there, providing a convenient marker of the untrustworthy parts of the plane, effectively demarcating the boundaries of the EoR window.  For these two reasons, Step 1 is unlikely to be an issue, at least not inside the EoR window.

More worrisome is Step 3, where the power spectrum bias of contaminants is subtracted off.  If we define ``contaminants" to be ``everything but the cosmological $21\,\textrm{cm}$ signal", there are two potential sources of bias: foregrounds and noise.  The subtraction of these biases is not followed by a downweighting analogous to the application $\mathbf{C}^{-1}$ in Step 1.  Moreover, whereas one could argue that the foreground bias is likely to be large only outside the EoR window, the noise bias will spread throughout the $k_\perp$-$k_\parallel$ plane.  This noise bias will also be quite large, as current experiments are firmly in the regime where the signal-to-noise is below unity.  It would therefore be advantageous to avoid bias subtractions altogether if possible.

To avoid having to subtract foreground bias, we simply redefine what we mean by contaminants/junk.  If we modify our mission to be one where we are measuring the power spectrum of total sky emission instead of the power spectrum of the cosmological $21\,\textrm{cm}$ signal, the foreground contribution to the bias term no longer exists, as foregrounds now count as part of the signal we wish to measure.  Of course, nothing has really changed, for we have simply ignored the subtraction of the foreground bias by redefining what we mean by ``contaminants".  The method is still optimal for measuring the power spectrum of the sky emission---though now it will not provide the absolute best possible limits on the EoR power spectrum.  Within the EoR window, this should result in little degradation of our final constraints, for in this region foreground contamination is expected to be negligible, and the power spectrum of the cosmological signal should be essentially identical to the power spectrum of total sky emission.  In any case, this is an assumption that can be checked in the final results, and represents a conservative assumption throughout Fourier space since foreground power is necessarily positive.  As detailed low-frequency foreground observations are conducted, it may be possible to achieve more sensitivity in foreground contaminated regions by taking advantage of more detailed maps and developing more faithful models. This task is left to future power spectrum estimation studies.

In contrast, escaping to the safe confines of the EoR window alone is not sufficient to eliminate the instrumental noise portion of the bias term, for the instrumental noise bias pervades the entire $k_\perp$-$k_\parallel$ plane.  To eliminate the noise bias, one can choose to compute not the auto-power spectrum of a single data cube with itself, but instead to compute the cross-power spectrum of two data cubes that are formed from data from interleaved (i.e., odd and even) time samples.  Since the instrumental noise is uncorrelated in time, this has the effect of automatically removing the instrumental noise bias\footnote{The reader may object to this by (correctly) pointing out that there exist errors that are correlated in time, with calibration errors being a prime example.  The result would be a cross-power spectrum that still retained a bias.  However, this does not invalidate the cross-power spectrum approach, in the following sense.  While biases will make our estimates of the power spectrum imperfect, these estimate will not be incorrect---the final (biased) power spectra will still represent perfectly rigorous upper limits on the cosmological power, provided we are conservative about how we estimate our error bars.  We will discuss how to make such conservative error estimates later on in this section and in Section \ref{sec:CovarModeling}.}.

More explicitly, we can form a bandpower estimate of the cross-power spectrum by simply computing
\begin{equation}
\label{crossEstimator}
\widehat{p}^ \textrm{cross}_\alpha = \mathbf{x}^t_1 \mathbf{E}^\alpha \mathbf{x}_2,
\end{equation}
where $\mathbf{x}_1$ and $\mathbf{x}_2$ are the data vectors for the two time inter-leaved data cubes, and for notational brevity we have defined $\mathbf{E}^\alpha \equiv \frac{1}{2} \sum_\beta M_{\alpha \beta} \mathbf{C}^{-1} \mathbf{C}_{,\beta} \mathbf{C}^{-1}$.  For notational cleanliness we will omit the $-\mathbf{m}$ term in our power spectrum estimator for this section only, with the understanding that $\x$ signifies the data vector \emph{after} the best-guess foreground model has already been subtracted.  In a similar fashion, $\x_\textrm{fg}$ refers to the foreground residuals, post-subtraction.

To see that the cross-power spectrum has no noise bias, let us decompose the data vectors $\mathbf{x}_i$ into the sum of $ \mathbf{s}$ and $\mathbf{n}_i$, the signal and noise components respectively, where the signal component has no index because it does not vary in time (note also that following the discussion above, any true sky emission counts as signal, so that $\mathbf{s} \equiv \mathbf{x}_{21} + \mathbf{x}_{\textrm{fg}}$).  Inserting this decomposition into the preceding equation and taking the expectation value of the result gives
\begin{align}
\langle \widehat{p}^ \textrm{cross}_\alpha \rangle =& \langle (\mathbf{s} + \mathbf{n}_1)^t \mathbf{E}^\alpha (\mathbf{s} + \mathbf{n}_2)\rangle \nonumber \\
=& \langle \mathbf{s}^t \mathbf{E}^\alpha \mathbf{s} \rangle + \langle \mathbf{n}_1 \rangle^t \mathbf{E}^\alpha \mathbf{s} \nonumber \\
& + \mathbf{s}^t \mathbf{E}^\alpha \langle \mathbf{n}_2 \rangle  + \langle \mathbf{n}_1 \mathbf{E}^\alpha \mathbf{n}_2 \rangle \nonumber \\
=& \langle \mathbf{s}^t \mathbf{E}^\alpha \mathbf{s} \rangle,
\end{align}
where the last equality holds because the instrumental noise has zero mean, i.e. $\langle \mathbf{n}_i \rangle =0$, and no cross-correlation between different times, i.e. $\langle \mathbf{n}_1 \mathbf{n}_2 \rangle = 0$.  The resulting estimator depends only on the power spectrum of the signal, and there is no additive bias.

Importantly, however, we emphasize that while we have eliminated noise \emph{bias} by computing a cross-power spectrum, we have not eliminated noise \emph{variance}.  In other words, the instrumental nosie will still contribute to the error bars.  To see this, consider the variance in our estimator, which is given by
\begin{align}
\boldsymbol \Sigma_{\alpha \beta}^{\textrm{cross}} =& \langle \widehat{p}^ \textrm{cross}_\alpha  \widehat{p}^ \textrm{cross}_\beta \rangle - \langle \widehat{p}^ \textrm{cross}_\alpha \rangle \langle \widehat{p}^ \textrm{cross}_\beta \rangle \nonumber \\ 
=& \langle  \mathbf{x}^t_1 \mathbf{E}^\alpha \mathbf{x}_2  \mathbf{x}^t_1 \mathbf{E}^\beta \mathbf{x}_2 \rangle - \langle  \mathbf{x}^t_1 \mathbf{E}^\alpha \mathbf{x}_2\rangle \langle  \mathbf{x}^t_1 \mathbf{E}^\beta \mathbf{x}_2 \rangle
\end{align}
The second term simplifies to
\begin{equation}
 \langle \widehat{p}^ \textrm{cross}_\alpha \rangle \langle \widehat{p}^ \textrm{cross}_\beta \rangle = \sum_{ijkl} \langle \mathbf{x}^i_1 \mathbf{x}^j_2 \rangle \langle \mathbf{x}^k_1 \mathbf{x}^l_2 \rangle \mathbf{E}_{ij}^\alpha \mathbf{E}_{kl}^\beta.
\end{equation}
Similarly, the first term is equal to
\begin{align}
 \langle \widehat{p}^ \textrm{cross}_\alpha  \widehat{p}^ \textrm{cross}_\beta \rangle = \sum_{ijkl} &\langle  \mathbf{x}^i_1 \mathbf{x}^j_2 \mathbf{x}^k_1 \mathbf{x}^l_2 \rangle \mathbf{E}_{ij}^\alpha \mathbf{E}_{kl}^\beta \nonumber \\
 = \sum_{ijkl}& \bigg(\langle  \mathbf{x}^i_1 \mathbf{x}^j_2 \rangle \langle \mathbf{x}^k_1 \mathbf{x}^l_2 \rangle + \langle  \mathbf{x}^i_1 \mathbf{x}^k_1 \rangle \langle \mathbf{x}^j_2 \mathbf{x}^l_2 \rangle \nonumber \\ 
&+ \langle  \mathbf{x}^i_1 \mathbf{x}^l_2 \rangle \langle \mathbf{x}^j_1 \mathbf{x}^k_2 \rangle\bigg) \mathbf{E}_{ij}^\alpha \mathbf{E}_{kl}^\beta,
\end{align}
where in the last equality we assumed Gaussian distributed data to simplify the four-point correlation.\footnote{In principle, $\x$ may exhibit departures from Gaussianity, since foregrounds are typically not Gaussian-distributed.  However, there are several reasons to expect deviations from non-Gaussianity to be unimportant.  First, the most flagrantly non-Gaussian foregrounds are typically those that are bright.  When we analyze real data in Section \ref{sec:WorkedExample}, we alleviate this problem by analyzing only a relatively clean part of the sky.  In addition, recall that in this section, $\x$ represents the data \emph{after} a best-guess model of foregrounds has been subtracted from the original measurements.  Thus, the crucial probability distribution to consider is not the foregrounds themselves, but rather the \emph{deviations} from the foregrounds, which are likely to be better-approximated by a Gaussian distribution.}  Our bandpower covariance is now
\begin{align}
\boldsymbol \Sigma_{\alpha \beta}^{\textrm{cross}} =  \sum_{ijkl} & \bigg( \langle  \mathbf{x}^i_1 \mathbf{x}^k_1 \rangle \langle \mathbf{x}^j_2 \mathbf{x}^l_2 \rangle \nonumber \\ 
&+ \langle  \mathbf{x}^i_1 \mathbf{x}^l_2 \rangle \langle \mathbf{x}^j_1 \mathbf{x}^k_2 \rangle\bigg) \mathbf{E}_{ij}^\alpha \mathbf{E}_{kl}^\beta.
\end{align}
The first term in this expression consists only of auto-correlations, which contain both noise and signal:
\begin{equation}
\langle \mathbf{x}_1 \mathbf{x}_{1}^t \rangle = \langle (\mathbf{s}+\mathbf{n}_1) (\mathbf{s}^t + \mathbf{n}_1^t) \rangle -\langle \mathbf{s}\rangle \langle \mathbf{s} \rangle^t= \mathbf{S} + \mathbf{N} = \mathbf{C},
\end{equation}
where we have defined $\mathbf{C}$ to be the total data covariance (as defined in Section \ref{sec:IdealObs}), $\mathbf{S} \equiv \langle \mathbf{s} \mathbf{s}^t \rangle-\langle \mathbf{s} \rangle \langle \mathbf{s} \rangle^t$ is the sky signal covariance (as per the discussion earlier in this section), and $\mathbf{N} \equiv \langle \mathbf{n}_1 \mathbf{n}_1^t \rangle = \langle \mathbf{n}_2 \mathbf{n}_2^t \rangle$ is the instrumental noise covariance.  We have assumed that there is no correlation\footnote{Note that this assumption has nothing to do with whether or not the instrument is sky-noise dominated.  A sky-noise dominated instrument will have instrumental noise whose \emph{amplitude} depends on the sky temperature, but the actual noise fluctuations will still be uncorrelated with the sky signal.} between the sky emission and the instrumental noise, so that $  \langle \mathbf{s} \mathbf{n}^t_1 \rangle = \langle \mathbf{s} \mathbf{n}^t_2 \rangle=0$.

The second term in our bandpower covariance consists only of cross-correlations, and thus contains no noise covariance:
\begin{equation}
\langle \mathbf{x}_1 \mathbf{x}_2^t \rangle = \langle (\mathbf{s}+\mathbf{n}_1) (\mathbf{s}^t + \mathbf{n}_2^t) \rangle = \mathbf{S}.
\end{equation}
Putting everything together, we obtain
\begin{equation}
\label{properCrossCovar}
\boldsymbol \Sigma_{\alpha \beta}^{\textrm{cross}}  = \textrm{tr}\! \left[ \mathbf{C} \mathbf{E}^\alpha \mathbf{C} \mathbf{E}^\beta \right]+ \textrm{tr}\! \left[ \mathbf{S} \mathbf{E}^\alpha \mathbf{S} \mathbf{E}^\beta \right].
\end{equation}
This, then, is the error covariance of our cross power spectrum estimator.  It gives less variance than the expression for the auto power spectrum, which in the notation of this section takes the form
\begin{equation}
\label{autoCovar}
\boldsymbol \Sigma_{\alpha \beta}^{\textrm{auto}}  = 2 \textrm{tr}\! \left[ \mathbf{C} \mathbf{E}^\alpha \mathbf{C} \mathbf{E}^\beta \right].
\end{equation}
Despite this difference between equations \ref{properCrossCovar} and \ref{autoCovar}, one may conservatively opt to use the above covariance matrix for the auto-power spectrum to estimate error bars even when using Equation \eqref{crossEstimator} to estimate the power spectrum itself.  In fact, it may be prudent to make this choice because there exists the possibility that the noise between interleaved time samples may not be truly uncorrelated, making the true errors closer to those described by $\boldsymbol \Sigma^{\textrm{auto}}$.  In our worked example with MWA data in Section \ref{sec:WorkedExample}, we will conservatively use Equation \eqref{autoCovar} to estimate the errors of our cross-power spectrum.  The task of characterizing the noise properties of the instrument thoroughly enough to eliminate this assumption is left to future work on a larger data set.

In summary, uncertainties in noise and foreground properties make it desirable to avoid trying to extract weak signals by performing subtractions between two large numbers (the contamination-dominated data and the possibly inaccurate contaminant models).  Mathematically, the greatest concern comes with the subtraction of the noise and foreground biases from power spectra estimates.  To deal with the residual noise bias, one may evaluate cross-power spectra between interleaved time samples rather than auto-power spectra.  To deal with the foreground bias, one can conservatively elect to simply leave it in when placing upper limits on the cosmological signal, and rely on the robustness of the EoR window to separate out the foregrounds from the cosmological $21\,\textrm{cm}$ signal.  In effect, one can practice foreground avoidance rather than foreground subtraction, since the former (if it is sufficient for a detection of the cosmological signal) will be more robust than the latter in the face of foreground uncertainties.  Finally, as a brute-force safeguard, to quantify such uncertainties, one can always vary the foreground model used in power spectrum estimation, as we do in Section \ref{sec:CovarModeling} when we apply our methods to the worked example of MWA data.

\subsection{A real-world obstacle: incomplete $uv$-coverage}
\label{sec:IncompleteUV}

While the methods of the previous section allow one to alleviate the effects of foreground modeling uncertainty, it is impossible to avoid the fact that real interferometers are imperfect imaging instruments.  This is because a real interferometer will inevitably have $uv$-coverage that is non-ideal in two ways.  First, the coverage is non-uniform, resulting in images that have been convolved with non-trivial synthesized beam kernels.  Second, the $uv$-coverage is incomplete, in that certain parts of the $uv$-plane are not sampled at all.  The idealized methods of Section \ref{sec:IdealObs} deals with neither problem, and in this section with augment the formalism to rectify this.

Assume for a moment that $uv$ coverage is complete (so that there are no ``holes" in the $uv$-plane), but not necessarily uniform.  In such a scenario, one has measured an unevenly weighted sample of the Fourier modes of the sky.  The effect of this non-trivial weighting needs to be accounted for when measuring the power spectrum, since $uv$ coordinates roughly map to $k_\perp$.  A failure to do so would therefore result in the final power spectrum estimate being multiplied by some function of $k_\perp$ corresponding to the $uv$ distribution.

Put another way, the $uv$ distribution of an interferometer defines its synthesized beam, the kernel with which the true sky has been convolved in the production of our image data cube.  The equations of Section \ref{sec:IdealObs} assume that this convolution has already been undone.  Thus, we must first perform this step, which in our notation may be written as
\begin{equation}
\label{eq:Deconv}
\mathbf{x} = \mathbf{B}^{-1} \mathbf{x}^\prime,
\end{equation}
where $\mathbf{x}^\prime$ represents the convolved data vector, $\mathbf{B}$ is the convolution matrix encoding the effects of the synthesized beam, and $\mathbf{x}$ is the processed data vector that is fed into Equation \eqref{eq:unnormBands}.  Note that this application of $\mathbf{B}^{-1}$ is meant to undo only the effects of the synthesized beam, not the primary beam.

The above method assumes that the matrix $\mathbf{B}$ is invertible.  In practice, this will likely not be the case as parts of the $uv$ plane will be missed by the interferometer, resulting in a singular $\mathbf{B}$ matrix.  In what follows, we will present two different ways to deal with this.  The first is to modify the equations of Section \ref{sec:IdealObs} so that they accept the convolved images (the ``dirty maps") as input.  Since all the statistical information relevant to the power spectrum are encoded in the covariance matrix, we simply have to make the replacement
\begin{equation}
\mathbf{C} \equiv \langle \mathbf{x} \mathbf{x}^t \rangle - \langle \mathbf{x} \rangle \langle \mathbf{x} \rangle^t  \longrightarrow \langle \mathbf{x}^\prime \mathbf{x}^{\prime \,t} \rangle - \langle \mathbf{x}^\prime \rangle \langle \mathbf{x}^\prime \rangle^t.
\end{equation}
This amounts to
\begin{equation}
\label{eq:BCB}
\mathbf{C} \longrightarrow \mathbf{B} \left( \langle \mathbf{x} \mathbf{x}^t \rangle - \langle \mathbf{x} \rangle \langle \mathbf{x} \rangle^t \right) \mathbf{B}^t = \mathbf{B} \mathbf{C} \mathbf{B}^t.
 \end{equation}
Of course, changing the covariance matrix also changes $\mathbf{C}_{,\alpha}$, and we must propagate this change.  Differentiating the preceding equation with respect to the bandpower $p_\alpha$ gives the substitution
\begin{equation}
\label{eq:BCBderiv}
\mathbf{C}_{,\alpha} \longrightarrow \mathbf{B} \mathbf{C}_{,\alpha} \mathbf{B}^t.
\end{equation}
Since $\mathbf{C}_{,\alpha}$ is the response of the data covariance $\mathbf{C}$ to the bandpower $p_\alpha$, this is simply a statement of the fact that if our data consists of dirty maps, the revised $\mathbf{C}_{,\alpha}$ matrix should encode the response of a dirty map's data covariance to the bandpower.  With the substitutions given by Equations \eqref{eq:BCB} and \eqref{eq:BCBderiv}, the rest of the equations of Section \ref{sec:IdealObs} can be used unchanged.  In the limit of an invertible $\mathbf{B}$ matrix, it is straightforward to show that this is equivalent to using Equation \eqref{eq:Deconv}.

The second method for dealing with a singular $\mathbf{B}$, which was proposed in Ref. \cite{DillonFast}, is to replace the ill-defined inverse matrix $\mathbf{B}^{-1}$ with a pseudoinverse given by
\begin{equation}
\boldsymbol \Pi \left( \mathbf{B} + \gamma \mathbf{U} \mathbf{U}^\dagger \right)^{-1} \boldsymbol \Pi,
\end{equation}
where $\gamma$ is a non-zero but otherwise arbitrary real number, and $\boldsymbol \Pi$ is a projection matrix given by
\begin{equation}
\boldsymbol \Pi \equiv \mathbf{I} - \mathbf{U} (\mathbf{U}^\dagger \mathbf{U})^{-1} \mathbf{U}^\dagger.
\end{equation}
The matrix $\mathbf{U}$ specifies which modes on the sky are missing in the data as a result of unobserved pixels on the $uv$-plane.  It is constructed by computing the responses (on the sky) of each unobserved $uv$ pixel individually and storing each response as a column of $\mathbf{U}$.  As an example, in the flat-sky approximation the $\mathbf{U}$ matrix would have a sinusoid in each column, corresponding to the fringes that would have been observed by the interferometer had data not been missing in a particular $uv$ pixel.  If these modes were present in the covariance model (which might be the case, for example, if the covariance were constructed by modeling data from a different interferometer with different $uv$ coverage), then the inverse covariance $\mathbf{C}^{-1}$ in our estimator needs to be similarly replaced with the pseudoinverse:
\begin{equation}
\boldsymbol \Pi \left( \mathbf{C} + \gamma \mathbf{U} \mathbf{U}^\dagger \right)^{-1} \boldsymbol \Pi.
\end{equation}
Importantly, the pseudoinverse can be quickly multiplied by a vector using the previously discussed conjugate gradient method. Its usage therefore does not sacrifice any of the speedups that were identified in Section \ref{sec:DataVolume} for dealing with large data volumes.

\subsection{A real-world obstacle: missing data from RFI}
\label{sec:RFI}
In any practical observation, the presence of narrowband RFI will mean that certain RFI-contaminated frequency channels will need to be flagged as outliers and omitted from a final power spectrum analysis.  The result, once again, is the presence of gaps in the data, only this time the missing modes are complete frequency channels.  However, the pseudoinverse formalism of the previous section is quite flexible in that modes of any form can be projected out of the analysis.  Thus, to correctly account for RFI-flagged data, one simply uses the pseudoinverse in exactly the same way as one does to account for missing $uv$ data.

\subsection{A real-world obstacle: foreground leakage into the EoR window}
\label{sec:decorr}
As Equation \eqref{eq:phatWp} showed, estimates of the power spectrum are not truly local, in the sense that every bandpower estimate $\widehat{p}_\alpha$ corresponds to a weighted average of the true power spectrum, with weights specified by the window functions.  \citet{LT11} showed that these window functions can be quite broad, particularly in regions with high foreground contamination.  There is thus the danger that foreground power could leak into the EoR window.  Because the foregrounds are so much brighter than the cosmological signal, even a small amount of leakage could compromise the cleanliness of the EoR window.

Fortunately, one can exert some control over the shape of the window functions\footnote{The term ``window function" should not be confused with the term ``EoR window".  The former refers to the weights that specify the linear combination of the true bandpowers that each bandpower estimate represents, as per Equation \eqref{eq:phatWp}.  The latter refers to the region on the $k_\perp$-$k_\parallel$ plane that naturally has very low levels of foreground contamination, as illustrated in Figure \ref{fig:EoRWindow}.} by making wise choices regarding the form of $\mathbf{M}$ in Equation \eqref{eq:normingBands}, which in turn gives the window functions via $\mathbf{W} = \mathbf{M} \mathbf{F}$.  As discussed above, $\mathbf{M}$ must be chosen such that the rows of $\mathbf{W}$ sum to unity.  Beyond that requirement, however, an infinite number of choices are admissible.  One choice would be $\mathbf{M} = \mathbf{F}^{-1}$, which gives $\mathbf{W} = \mathbf{I}$ (i.e., delta function windows).  This would certainly minimize the amount of leakage into the EoR window, but it comes at a high price: the resulting error bars on the power spectrum measurement---the diagonal elements of $\boldsymbol \Sigma$ from Equation \eqref{eq:CovP}---tend to be large, reflecting the data's inability to make highly localized measurements in Fourier space when the survey volume is finite.

On the other extreme, the error bars predicted by $\boldsymbol \Sigma$ can be shown to be their smallest possible if $\mathbf{M}$ is taken to be diagonal \citep{THX2df}.  However, this gives broader window functions, for it is via the smoothing/binning effect of these broad window functions that the small errors can be achieved.  One can also argue that the level of smoothing dictated by this approach is excessive, since the resulting bandpowers have positively correlated errors.  (To see this, note that up to a row-dependent normalization, the error covariance matrix takes the form $\boldsymbol \Sigma \sim \mathbf{F}$.  Since all elements of a Fisher matrix must necessarily be non-negative, this implies that all cross-covariances of the estimated bandpowers have positively correlated errors unless $\mathbf{F}$ is diagonal, which is rarely the case).

As a compromise option, we advise $\mathbf{M} \sim \mathbf{F}^{-1/2}$ (again after a normalization of each row so that the window functions sum to unity).  This choice gives window functions that are narrower than those for a diagonal $\mathbf{M}$ while maintaining reasonably small error bars.  In addition, an inspection of Equation \eqref{eq:CovP} reveals that this method gives a diagonal $\boldsymbol \Sigma$, which means that errors between different bandpowers are uncorrelated.

In Section \ref{sec:MethodDemo}, we use MWA data to demonstrate the crucial role that the $\mathbf{M} \sim \mathbf{F}^{-1/2}$ choice plays in preserving the cleanliness of the EoR window.\footnote{Of course, there exist other choices that are more elaborate than the three considered in this paper.  For example, with exquisite foreground and instrumental modeling, one could imagine first decorrelating to delta-function windows by setting $\M = \F^{-1}$ in an attempt to ``perfectly" contain the foregrounds to regions outside the EoR window, and then to re-smooth the bandpowers within the window to reduce the variance.  This is a promising avenue for future investigation, but for this paper our goal is simply to apply the $\F^{-1/2}$ decorrelator to real data (see Section \ref{sec:MethodDemo}) to demonstrate the feasibility of containing foregrounds using decorrelation techniques.}

\subsection{A real-world obstacle: ensuring that binning doesn't destroy error properties}
\label{sec:cylindToSph}
 
In previous sections, we have discussed how one can preserve all the desirable properties of the power spectrum estimator of Section \ref{sec:IdealObs} in the face of all the real-world complications presented in Sections \ref{sec:DataVolume} through \ref{sec:decorr}.  The result is a rigorous yet practical estimator for the cylindrical power spectrum $P_\textrm{cyl} (k_\perp, k_\parallel)$.  We now turn to the problem of binning the cylindrical power spectrum into the cosmologically relevant spherical power spectrum $P_\textrm{sph} (k)$, with a special emphasis on the preservation of the information content of our estimator.

Just as with the cylindrical power spectrum, we parameterize the spherical power spectrum as piecewise constant, so that all the information is encoded in a vector of bandpowers $\mathbf{p}^{\textrm{sph}}$, so that:
\begin{equation}
p_\alpha^\textrm{sph} \equiv P_{\textrm{sph}} (k^\alpha).
\end{equation}
The spherical bandpowers are related to estimates of the cylindrical bandpowers $\widehat{\mathbf{p}}^\textrm{cyl}$ by the equation
\begin{equation}
\widehat{\mathbf{p}}^\textrm{cyl} = \mathbf{A} \mathbf{p}^\textrm{sph} + \boldsymbol \varepsilon,
\end{equation}
where $\mathbf{A}$ is a matrix of size $N_\textrm{cyl} \times N_\textrm{sph}$ of 1s and 0s that relates $k_\perp$-$k_\parallel$ pairs to $k$ bins, with $N_\textrm{cyl}$ and $N_\textrm{sph}$ equal to the number of cells in the $k_\perp$-$k_\parallel$ plane and the number of spherical $k$ bins respectively.  The vector $\boldsymbol \varepsilon$ is a random vector of errors on $\widehat{\mathbf{p}}^\textrm{cyl}$.  It has zero mean (assuming that one has taken the care to avoid additive bias in our estimator of the cylindrical bandpowers, as discussed above), but non-zero covariance equal to $\boldsymbol \Sigma^\textrm{cyl} \equiv \langle \boldsymbol \varepsilon \boldsymbol \varepsilon^t \rangle$, where $\boldsymbol \Sigma^\textrm{cyl}$ is given by either Equation \eqref{properCrossCovar} or \eqref{autoCovar}, depending on whether the cylindrical bandpowers were computed using cross or auto-power spectra.  (The methods presented in this section are applicable either way).
 
Our goal is to construct an optimal, unbiased estimator of $\mathbf{p}^\textrm{sph}$ from $\widehat{\mathbf{p}}^\textrm{cyl}$.  This is a solved problem \citep{TegmarkCMBmapsWOLosingInfo}, and the best estimator $\widehat{\mathbf{p}}^\textrm{sph}$ is given by
\begin{equation}
\label{eq:Cylind2SphEst}
\widehat{\mathbf{p}}^\textrm{sph} = [ \mathbf{A}^t \boldsymbol \Sigma_\textrm{cyl}^{-1} \mathbf{A} ]^{-1} \mathbf{A}^t \boldsymbol \Sigma_\textrm{cyl}^{-1} \widehat{\mathbf{p}}^\textrm{cyl},
\end{equation}
with the final error covariance on the spherical bandpowers given by
\begin{equation}
\label{eq:Cylind2SphCovar}
\boldsymbol \Sigma^\textrm{sph}_{\alpha \beta} \equiv \langle \widehat{\mathbf{p}}^\textrm{sph}_\alpha \widehat{\mathbf{p}}^\textrm{sph}_\beta \rangle -  \langle \widehat{\mathbf{p}}^\textrm{sph}_\alpha \rangle \langle  \widehat{\mathbf{p}}^\textrm{sph}_\beta \rangle =  [ \mathbf{A}^t \boldsymbol \Sigma_\textrm{cyl}^{-1} \mathbf{A} ]^{-1} .
\end{equation}
Since the $\mathbf{A}$ matrix has (by construction) a single 1 per row and zeros everywhere else, an inspection of Equation \eqref{eq:Cylind2SphCovar} reveals that a diagonal $\boldsymbol \Sigma^{\textrm{cyl}}$ implies a diagonal $\boldsymbol \Sigma^\textrm{sph}$.  In other words, the estimator given by Equation \eqref{eq:Cylind2SphEst} preserves the decorrelated nature of the $\M \sim \F^{-1/2}$ version of the cylindrical power spectrum estimator defined in Section \ref{sec:decorr}.  This will not be the case for an arbitrary estimator (such as one that is formed from taking uniformly weighted Fast Fourier Transforms, then squaring and binning).  We also emphasize that if one does not choose to use decorrelated cylindrical bandpower vectors, Equations \eqref{eq:Cylind2SphEst} and \eqref{eq:Cylind2SphCovar} require that one keep full track of the off-diagonal terms of $\boldsymbol \Sigma_\textrm{cyl}^{-1}$.  Without it, a consistent propagation of errors to the spherical power spectrum is not possible, and may even lead to a mistakenly claimed detection of the cosmological signal, as we discuss in Section \ref{sec:MethodDemo} and in Appendix \ref{sec:ErrorCovAppendix}.

Just as with the cylindrical power spectra, we would like to compute the window functions.  The definition of the spherical window functions are exactly analogous to that provided in Equation \eqref{eq:phatWp} for the cylindrical power spectrum, so that
\begin{equation}
\langle \mathbf{\widehat{p}}^\textrm{sph} \rangle = \mathbf{W}^\textrm{sph} \mathbf{p}^\textrm{sph}.
\end{equation}
Taking the expectation value of Equation \eqref{eq:Cylind2SphEst}, we have
\begin{align}
\langle \mathbf{\widehat{p}}^\textrm{sph} \rangle &= [ \mathbf{A}^t \boldsymbol \Sigma_\textrm{cyl}^{-1} \mathbf{A} ]^{-1} \mathbf{A}^t \boldsymbol \Sigma_\textrm{cyl}^{-1} \langle \widehat{\mathbf{p}}^\textrm{cyl} \rangle \nonumber \\
 &= [ \mathbf{A}^t \boldsymbol \Sigma_\textrm{cyl}^{-1} \mathbf{A} ]^{-1} \mathbf{A}^t \boldsymbol \Sigma_\textrm{cyl}^{-1}  \mathbf{W}^\textrm{cyl} \mathbf{A} \mathbf{p}^{\textrm{sph}},
\end{align}
where we have used the definition of the cylindrical window functions to say that $\langle \widehat{\mathbf{p}}^\textrm{cyl} \rangle = \mathbf{W} \mathbf{p}^\textrm{cyl}$, as well as the fact that $\mathbf{p}^\textrm{cyl} = \mathbf{A} \mathbf{p}^\textrm{sph}$ (with no error term because we are relating the true cylindrical bandpowers to the true spherical bandpowers).  Inspecting this equation, we see that
\begin{equation}
\mathbf{W}^\textrm{sph} = [ \mathbf{A}^t \boldsymbol \Sigma_\textrm{cyl}^{-1} \mathbf{A} ]^{-1} \mathbf{A}^t \boldsymbol \Sigma_\textrm{cyl}^{-1}  \mathbf{W}^\textrm{cyl} \mathbf{A}.
\end{equation}
Therefore, by measuring the width of the spherical window functions (rows of $\mathbf{W}^\textrm{sph}$), one can place rigorous horizontal error bars on the final spherical power spectrum estimate.

\subsection{Summary of the issues}
In the last few sections, we have provided techniques for dealing with a number of real-world obstacles.  These include:
\begin{enumerate}
\item Taking advantage of the flat-sky approximation and the rectilinearity of data cubes, as well as the conjugate gradient algorithm for matrix inversion to allow large data sets to be analyzed quickly.
\item Using cross-power spectra rather than auto-power spectra in order to eliminate noise bias.
\item Replacing inverses with pseudoinverses to deal with data that has missing spatial modes (due to incomplete $uv$ coverage) and missing frequency channels (due to RFI).
\item Performing power spectrum decorrelation to avoid the leakage of foreground power into the EoR window.
\item Binning of cylindrical power spectra into spherical power spectra in a way that preserves desirable error properties.
\end{enumerate}
Crucial to this is the fact that these techniques all operate under a self-consistent framework.  This allows faithful error propagation that accurately captures how various real-world effects act together.  For example, it was shown in \cite{DillonFast} that properly accounting for pixelization effects in Equation \eqref{eq:pixDef} results in low Fisher information at high $k_\parallel$, providing a marker for parts of the $k_\perp$-$k_\parallel$ plane that cannot be well-constrained because of finite spectral resolution.  The identification of such a region would be trivial if one had spectrally contiguous data, for then one would simply say that the largest measurable $k_\parallel$ was roughly $1/\Delta L_\parallel$, where $\Delta L_\parallel$ is the width of a single frequency channel mapped into a cosmological line-of-sight distance.  However, such a straightforward analysis no longer applies when there are RFI gaps in the data at arbitrary locations.  In contrast, the unified framework presented in this paper allows all such complications to be folded in correctly.

\section{A worked example: early MWA data}
\label{sec:WorkedExample}
Now that we have bridged the gap between theoretical techniques for analyzing ideal data and the numerous challenges presented by real data, we are ready to bring together our methods, specify a covariance model, and estimate power spectra from MWA 32-tile prototype (MWA-32T) data.  The data were taken between the 21st and 29th of March 2010, the first observing campaign during which data were taken that were scientifically useful.  The observations are described in more detail by \cite{ChrisMWA}.  Real data affords us two opportunities.  In this section, we look at the data to examine and quantify the differences between power spectrum estimators and the pitfalls associated with choice of estimator.  In Section \ref{sec:earlyResults}, we take advantage of everything we have developed to arrive at interesting new foreground results and a limit on the 21 cm brightness temperature power spectrum.

\subsection{Description of observations}

All of the data used for this paper were taken on the MWA-32T system.  This system has since been upgraded to a 128-tile instrument (MWA-128T; \citet{TingaySummary, BowmanMWAScience}), but in this paper we focus exclusively on MWA-32T data, reserving the MWA-128T data for future work.

The MWA-32T instrument consisted of 32 phased-array ``antenna tiles" which served as the primary collecting elements.  Each tile contained 16 dual linear-polarization wideband dipole antennas which were combined to form a steerable beam with a full width at half maximum (FWHM) size of $\sim25^\circ$ at 150~MHz.  The array had an approximately circular layout with a maximum baseline length of $\sim340$~m, and a minimum baseline length of 6.6~m, although the shortest operating baseline during this observational campaign was 16~m.  After digitization, filtering, and correlation, the final visibilities had a 1~second time resolution and 40~kHz spectral resolution over a 30.72~MHz bandwidth.    The instrumental capabilities are summarized in Table~\ref{tab:mwa32t}.

\begin{table*}
\begin{center}
\begin{tabular}{ll}
\hline
Field of View (Primary Beam Width) & $\sim25^\circ$ at 150~MHz \\
Angular Resolution & $\sim 20'$ at 150~MHz \\
Collecting Area & $\sim 690~{\rm m}^2$ towards zenith at 150~MHz  \\
Polarization & Linear X-Y \\
Frequency Range & 80~MHz to 300~MHz \\
Instantaneous Bandwidth & 30.72~MHz \\
Spectral Resolution & 40~kHz \\
\hline
\end{tabular}
\caption{MWA-32 Instrument Parameters\label{tab:mwa32t}}
\end{center}
\end{table*}

For our worked example, we concentrate on March 2010 observations of the MWA ``EoR2'' field.  It is centered located at ${\rm R.A.(J2000)} = 10^{\rm h}\ 20^{\rm m}\ 0^{\rm s}$,
${\rm decl.(J2000)} = -10^\circ\ 0'\ 0''$, and is one of two fields at high Galactic latitude that have been identified by the MWA collaboration as candidates for deep integrations, owing to their low brightness temperature in low frequency measurements of Galactic emission \citep{Haslam,Angelica}.  For further details about the observational campaign or the EoR2 field, please see \citet{ChrisMWA}, which was based on the same set of observations as the ones used in this paper.  

Observations covered three 30.72~MHz wide bands, centered at $123.52~{\rm MHz}$, $154.24~{\rm MHz}$ and $184.96~{\rm MHz}$, corresponding to a redshift range of $6.1 < z < 12.1$ (the redshift range of the results presented in this work is slightly smaller because of data flagging)  for the 21~cm signal.  The $123.52~{\rm MHz}$ and $154.24~{\rm MHz}$ bands were observed for approximately 5 hours each, and the $184.96~{\rm MHz}$ band was observed for approximately 12 hours.

These early data from the prototype have provided us with a set of test data that enabled development of extensive analysis methods and software on which the results of this paper are based.  The early prototype had shortcomings (e.g., mismatched cables, receiver firmware errors, correlator timing errors) that compromised the calibration to some extent, raising the apparent noise level.  Additionally, the instrument was only operating with $\lesssim 29$ tiles, and with a 50\% duty cycle throughout the course of these observations.  We account for this in Section \ref{sec:CovarModeling} by determining the magnitude of the noise empirically, in order to be able to place rigorously conservative upper limits on the cosmological power spectrum.  We expect that data from later prototype campaigns and from the full array will produce result closer to theoretical expectations.

\subsection{Mapmaking}
\label{sec:Mapmaking}
Before the data can be used as a worked example for our power spectrum estimator, however, we must convert the measured visibilities into a data cube of sky images at every frequency in our band. In other words, we must form the data vector $\mathbf{x}$, defined by Equation \eqref{eq:pixDef}, which serves as the input for our power spectrum pipeline.

To form the data vector, we performed the following steps.  First, we performed a reduction procedure similar to that described in \citet{ChrisMWA} for the initial flagging and calibration of the data.  Hydra~A was identified as the dominant bright source in the field, and used for calibration assuming a point source model.  The Hydra~A source model was then subtracted from the {\em uv} data.  As this same source model was also used for gain and phase calibration, this can be thought of as a ``peeling'' source removal procedure \citep{Noordam2004,vanderTol2007,Mitchell2008,Intema2009} on a single source.  Alternatively, in the absence of gridding artifacts, this is equivalent to imaging the point-source model and subtracting it from the data as part of the direct foreground subtraction step discussed in the first step of Section \ref{sec:crossPower} \citep{TegmarkCMBmapsWOLosingInfo}.

The subtracted data were imaged using the CASA task {\tt clean} without deconvolution to produce ``dirty'' images.  No multi-frequency synthesis was performed, so that the full 40~kHz spectral resolution of the data would be available.  The visibilities were gridded using w-projection kernels \citep{CornwellWProj} with natural (inverse-variance) weighting to produce maps at each frequency with a cell size of $3'$ over a $25.6^\circ$ field of view.  The resulting cubes contained $\sim200$ million voxels, with 512 elements along each spatial dimension and 768 elements in the frequency domain.  It is important to note that the pre-flagging performed on the data resulted in the flagging of entire frequency bands (which means that there are gaps in the final data cube).  Cubes were generated  for each 5~minute snapshot image.

The individual snapshot data cubes were combined using the primary beam inverse-variance weighting method described in \citet{ChrisMWA}.  The weighting and primary beams were simulated separately for each 40~kHz frequency channel in each 5~minute snapshot.  The combined maps and weights were saved, along with the effective point spread function at the center of the field.  Two additional data cubes were created by averaging alternating 5~minute snapshots (i.e. even numbered snapshots were averaged into one cube, and odd numbered snapshots were averaged into the other) so that they were generated from independent data, but with essentially the same sky and {\em uv} coverage properties.

A further flux scale calibration of the integrated cubes was performed using three bright point sources: MRC~1002-215, PG~1048-090, and PKS~1028-09 to set the flux scale on a channel-by-channel basis.  A two dimensional Gaussian fitting procedure was used to fit the peak flux of each of these sources in each 40~kHz channel of the data cube. Predictions for each source were derived by fitting a power law to source measurements from the 4.85~GHz Parkes-MIT-NRAO survey \citep{Griffith1995}, the 408~MHz Molonglo Reference Catalog \citep{Large1981}, the 365~MHz Texas Survey \citep{Douglas1996}, the 160~MHz and 80~MHz Culgoora Source List \citep{Slee1995} and the 74~MHz VLA Low-frequency Sky Survey \citep{Cohen2007}.  A weighted least-squares fit was then performed to calculate and apply a frequency-dependent flux scaling for the cube to minimize the square deviations of the source measurements from the power law models. 

An additional flagging of spectral channels was performed based on the root-mean-square (RMS) noise in each spectral channel of the cube.  A smooth noise model was determined by median filtering the RMS channel noise as a function of frequency (bins of 16 channels were used in the filtering).  Any channel with $5\sigma$ or larger deviations from the smoothed noise model was flagged.  Upon inspection, these additional flagged channels were observed to be primarily located at the edges of the coarse digital filterbank channels, which were corrupted due to an error in the receiver firmware.  After this procedure, approximately one third of the spectral channels were found to have been flagged.

Each individual map covered $25.6^\circ \times 25.6^\circ$ at a resolution of $3'$ with 768 frequency channels (40~kHz frequency resolution).  To decrease the computational burden of the covariance estimation, each map was subdivided into 9 subfields, and the pixels were averaged to a size of $15'$.  The data cubes were mapped to comoving cosmological coordinates using WMAP-7 derived cosmological parameters, with $\Omega_{\rm M}=0.266$, $\Omega_\Lambda=0.734$, $H_0=71~{\rm km}~{\rm s}^{-1}~{\rm Mpc}^{-1}$, and $\Omega_k \equiv 0$ \citep{WMAP7Cosmology}.  

At this point, the data cubes were ready to be used as input data to our power spectrum estimator, i.e., we had arrived at the final form of the data vector $\mathbf{x}$.  However, estimating power spectra and error statistics using the formalism of Section \ref{sec:Methods} also requires a covariance model, which we construct in the next section.

\subsection{Covariance model}
\label{sec:CovarModeling} 
We follow \cite{LT11} and \cite{DillonFast} in modeling the covariance matrix $\C$ as the sum of independent parts attributable to noise and foregrounds.  We leave off the signal covariance because it only contributes to the final error bars by accounting for cosmic variance---a completely negligible effect in comparison to foreground and noise-induced errors.  We adopt a conservative model of the extragalactic foregrounds by treating them as a Poisson random field of sources with fluxes less than 100 Jy, after the manual removal of Hydra A.  By treating all extragalactic foregrounds as ``unresolved," we effectively throw out information about which lines of sight are most contaminated by bright foregrounds.  As \cite{DillonFast} showed, future analyses can improve on our limits by including more information about the foregrounds. We begin with the parameterized covariance model of \cite{LT11}, 
\begin{align}
C_{ij}^\text{unresolved} =& \left(1.4 \times 10^{-3} \text{ }\frac{\text{K}}{\text{Jy}} \right)^2 \left( \int_0^{S_\text{cut}} S^2 \frac{dn}{dS}dS \right) \times \nonumber \\ 
&   \left(\frac{\nu_i \nu_j}{\nu_*^2} \right)^{-2-\bar{\kappa}} \exp\left[ \frac{\sigma_{\bar{\kappa}}^2}{2} \left(\ln \left[\frac{\nu_i \nu_j}{\nu_*^2} \right] \right)^2 \right] \times  \nonumber \\ 
& \exp \left[ \frac{(\mathbf{r}_{\perp i} - \mathbf{r}_{\perp j})^2}{2 \sigma_\perp^2}\right] \bigg(\Omega_\text{pix}\bigg)^{-1}
\label{eq:Umodel}
\end{align}
where $\nu_* = 150 \,\textrm{MHz}$ is a reference frequency, $\nu_i$ is the frequency of the $i$th voxel, which has an angular distance of $\mathbf{r}_{\perp i}$ from the field center.  The spectral index is $\bar{\kappa} = 0.5$, the uncertainty in the spectral index is $\sigma_\kappa = 0.5$, the clustering correlation length is $\sigma_\perp = 7'$, $\Omega_\text{pix}$ is the angular size of each pixel, the flux cut $S_\text{cut} = 100$ Jy, and $dn/dS$ is the differential source count from \cite{dimatteo1},
\begin{align}
\frac{dn}{dS} =& (4000 \text{ Jy}^{-1} \text{sr}^{-1})\nonumber \\ &\times \begin{cases} \left(\frac{S}{0.880 \text{ Jy}}\right)^{-2.51} & \text{for S $>$ 0.880 Jy} \\ \left(\frac{S}{0.880 \text{ Jy}}\right)^{-1.75} & \text{for S $\leq$ 0.880 Jy.} \end{cases}
\end{align}
We adapt this model for the fast power spectrum estimation method outlined in Section \ref{sec:DataVolume} by calculating the translationally invariant approximation to this model in the manner described in \cite{DillonFast}.

For the Galactic synchrotron, we also follow \cite{LT11} and \cite{DillonFast} for the parameterization of the synchrotron emission covariance.  Namely, we adopt $\bar{\kappa} = 0.8$, $\sigma_\kappa = 0.1$, $\sigma_\perp = 30^\circ$, and replace the first three terms of the covariance in Equation \ref{eq:Umodel} with $T^2_\text{synch} = (335.4 \text{ K})^2$.

Our model for the instrumental noise is adopted from \cite{DillonFast}, with one key difference: the overall normalization.  For each subband, we let the noise covariance matrix scale by a free multiplicative constant.  This is equivalent to treating the combination $T_\text{sys}^2 / ( A_\text{ant}^2 t_\text{obs})$ as a free parameter.  We then fit for that parameter by requiring the RMS difference between the two time slices---which should be free of sky signal---for the densely sampled inner region of $uv$ space and rescaling our noise covariance matrix to match.  The spatial structure of the covariance was left unchanged.  Even though the data is somewhat nosier than suggested by a first principles calculation assuming fiducial values for system temperature and antenna effective area, this empirical renormalization allows for an honest account of the errors introduced by instrumental effects.

To verify that our parameterization of the foregrounds was reasonable, we varied these parameters over an order of magnitude and found that they had little effect on our final power spectrum estimates, except at the lowest values of $k$.  There are two reasons for this: first, since we are only measuring the power spectrum of the sky, we need not worry about precisely subtracting foregrounds.  Second, because the noise in our instrument is still more than two orders of magnitude from the cosmological signal, in the EoR window our band power measurements will be noise dominated and agnostic to our foreground model.  Future analyses might include a more thorough treatment of the foregrounds, especially by utilizing the full power of the \citet{DillonFast} method to include information about the positions, fluxes, and spectral indices of individual point sources.

\subsection{Evaluating power spectrum estimator choices}
\label{sec:MethodDemo}

With both a data vector $\mathbf{x}$ and a covariance matrix $\mathbf{C}$ in hand, we can now apply the methods of Section \ref{sec:Methods} to estimate power spectra.  In doing so, we deal with real-world obstacles using all of the techniques that we have developed.  In this section, we show why all this is necessary.

In Section \ref{sec:decorr} we touted the choice of power spectrum estimator $\widehat{\mathbf{p}} = \M \mathbf{q}$ with $\M \sim \F^{-1/2}$ as a compromise solution between the choice with the smallest error bars, $\M \sim \Eye$, and the choice with the narrowest window functions, $\M \sim \F^{-1}$.  In the race to detect the power spectrum from the EoR, one might be tempted to aggressively seek out the smallest possible errors.  This could prove a deleterious choice, as we will now show using MWA-32T data.
 
\begin{figure*}[!ht] 
	\centering 
	\includegraphics[width=1\textwidth]{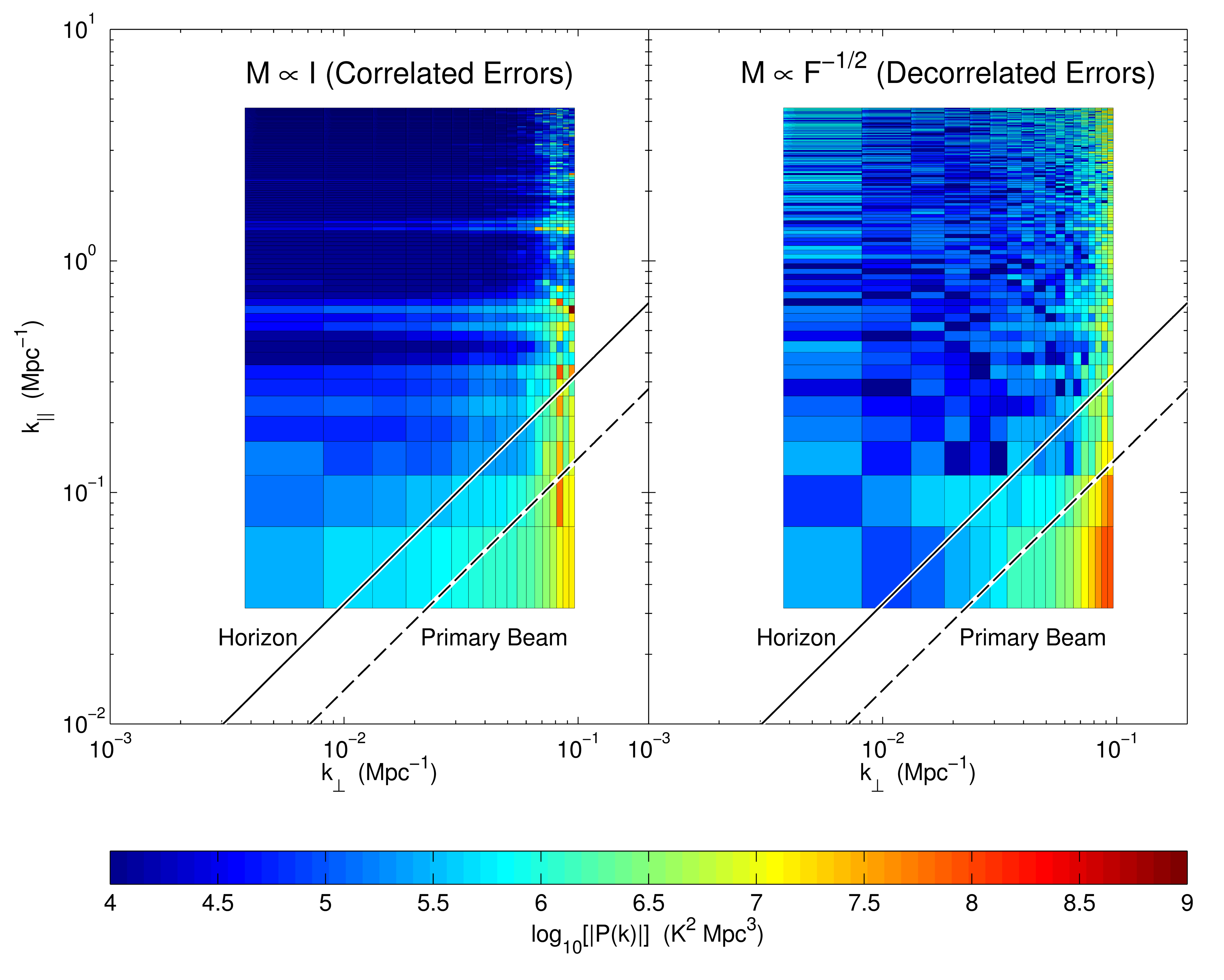}
	\caption{Unless one chooses a power spectrum estimator with decorrelated errors, foregrounds and other instrumental effects can leak significantly into the EoR window. Here we show the \textit{absolute value} of the cylindrical power spectrum estimate from the subband centered on 158 MHz ($z = 8.0$)  and averaged over all 9 fields.  On the left, we have set $\M \sim \Eye$.  On the right, $\M \sim \F^{-1/2}$.  We expect contamination from smooth spectrum foregrounds interacting with the chromatic synthesized beam to occupy the ``wedge" portion of Fourier space, defined in Equation \eqref{eq:wedge}.  Optimistically, the wedge is delimited by the extent of the main lobe of the primary beam; conservatively, we should not see bright foreground contamination beyond the horizon.  In the regions where the power spectrum is noise dominated, we expect little structure in the $k_\|$ direction in the EoR window above some moderate value of $k_\|$.  In the left panel, we see considerably more $k_\|$ structure in the form of horizontal bands, attributable to foreground contamination and instrumental effects, that has leaked into the putative EoR window.}
	\label{fig:2DPkComparison}
\end{figure*}

First, in Figure \ref{fig:2DPkComparison} we compare cylindrical power spectra, $\widehat{\mathbf{p}}$, generated using two different estimators of the power spectrum that we presented in Section \ref{sec:decorr}.\footnote{In our comparison of choices for $\M$, we drop the $\M \sim \F^{-1}$, $\delta$-function windows choice.  In addition to proving the noisiest estimator, it suffers from strong anti-correlated errors.  We adopt the perspective that the important comparison is between the ``obvious" choice, the minimum variance $\M \sim \Eye$, and our preferred choice with decorrelated errors, $\M \sim \F^{-1/2}$.}
On the left, we have used $\M \sim \Eye$, the estimator with the smallest error bars, and on the right we have used $\M \sim \F^{-1/2}$, the estimator with decorrelated errors.  In both cases, we have plotted the absolute value of the power spectrum estimates (which can be negative because they are cross-power spectra).  Because the two estimates are related to one another by an invertible matrix, they contain the same cosmological information. In a sense, the $\M \sim \F^{-1/2}$ method is the most honest estimator of the power spectrum because the band powers form a mutually exclusive and collectively exhaustive set of measurements.
In other words, they represent all the all the power spectrum information from the data, divided into independent pieces. 

Moreover, just because two sets of estimators have the same information content does not mean that they are equally useful for distinguishing the cosmological power spectrum from foreground contamination.   In Figure \ref{fig:2DPkComparison}, the minimum variance estimator for the power spectrum introduces considerable foreground contamination into the EoR window, demarcated by the expected angular extent of the wedge feature (which we introduced in Section \ref{sec:Intro} and will discuss in greater detail in Section \ref{sec:WedgeResults}).  Even highly suspect features at high $k_\perp$ where $uv$ coverage is spottiest seem to get smeared across $k_\perp$ and into the EoR window.  We cannot simply cut out the wedge from our cylindrical-to-spherical binning and expect a clean measurement of the power spectrum in the EoR window.
 
\begin{figure*}[!ht]  
	\centering 
	\includegraphics[width=1\textwidth]{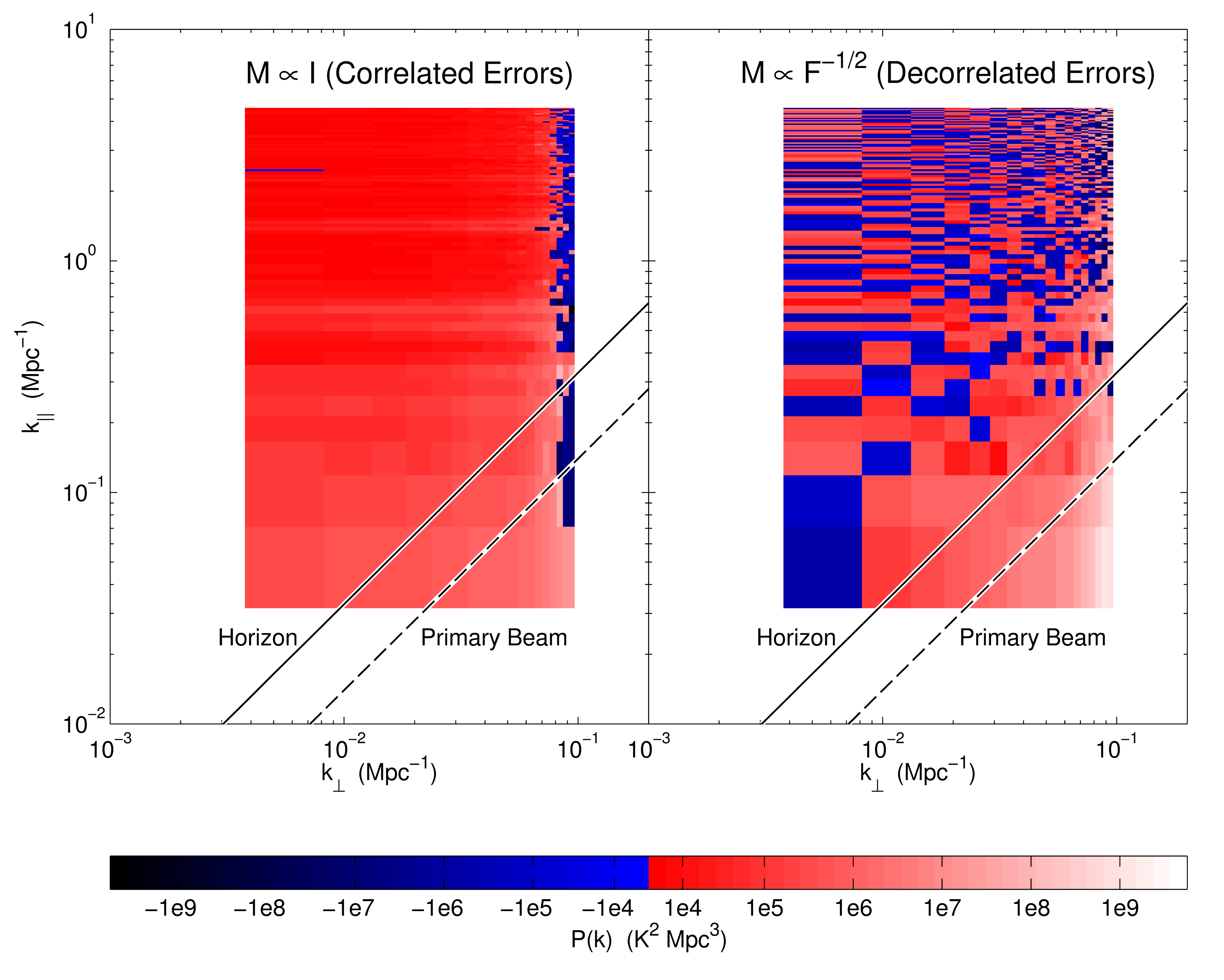}
	\caption{One advantage of calculating the cross power spectrum of interleaved time-slices of data is that we can easily tell which regions of Fourier space are noise dominated. Here we reproduce the power spectra from Figure \ref{fig:2DPkComparison} without taking the absolute value of $P(k)$.  By plotting with a discontinuous, sinh$^{-1}$ color scale, it is easy to see that the EoR window for our decorrelated power spectrum estimate (right panel) has roughly an equal number of positive and negative band power estimates---exactly what we would expect from a noise dominated region.  By contrast, our power spectrum estimate with correlated errors (left panel) shows positive power over almost all of Fourier space, indicating ubiquitous leakage of contaminants into the EoR window.}
	\label{fig:2DArcsinhPkComparison}
\end{figure*}

Looking closely at Figure \ref{fig:2DPkComparison}, one might notice that some regions of the EoR window on the lefthand panel still seem very clean---cleaner perhaps that the same regions in the righthand panel.  To examine that apparent fact, we plot $\widehat{p}_\alpha$ instead of $|\widehat{p}_\alpha|$ in Figure \ref{fig:2DArcsinhPkComparison}.  
To make the figure more intelligible, we have plotted colors based on an sinh$^{-1}$ color scale with a sharp color division at 0.  The sinh$^{-1}$ has the advantage of behaving linearly at small values of $\widehat{p}_\alpha$ and logarithmically at large positive or negative $\widehat{p}_\alpha$. 

What emerges is a striking difference between the two estimators.  For the reasons discussed in Section \ref{sec:crossPower}, we have chosen to estimate the cross power spectrum between two time-interleaved sets of observations.  As a result, we expect that instrumental noise should be equally likely to contribute positive power as it is to contribute negative power.  In noise dominated regions of the $k_\perp$-$k_\|$ plane, we expect about half of our measurements to be positive and about half to be negative.  That is exactly what we see in the EoR window of the $\M \sim \F^{-1/2}$ estimator.  However, the $\M \sim \Eye$ estimator in the lefthand panel clearly shows positive power throughout the entire supposed EoR window.  Though the magnitude of that power is not enormous---often it is well within the vertical error bars---the overall bias towards positive cross power means that sky signal is contaminating the EoR window.  This is precisely the problem we were worried about in Section \ref{sec:decorr} and the data have clearly manifested it.\footnote{Of course, as we noted in Section \ref{sec:decorr}, the choice of $\M \sim \F^{-1/2}$ is not unique in its ability to mitigate foreground leakage, and other choices certainly warrant future investigation.  Picking $\M \sim \F^{-1/2}$ is, however, a good choice for a first attempt at decorrelation, particularly given its various other desirable properties that we have described.  The important point here is that while $\M \sim \F^{-1/2}$ may not be necessarily optimal for containing foregrounds within the wedge, our results show that it is a reasonable one.  In contrast, the ``straightforward" approach of normalizing the power spectrum with the diagonal choice $\M \sim \Eye$ is clearly ill-advised.}

This also explains why there appeared to be less power in the EoR window of the lefthand panel of Figure \ref{fig:2DPkComparison}; by taking the absolute value of the weighted average of positive and negative quantities, we expect to measure a smaller absolute value of the power.  However, as this figure clearly shows, that weighted average is biased by foreground leakage.  And, even though there still appears to be a region just inside the EoR window that retains positive band powers consistent with foregrounds, that small amount of leakage can be attributed to finite sized windows functions and to calibration uncertainties.  Regardless, it does not appear to be an insurmountable limitation to the cleanliness of the EoR window; rather, it suggests that we should be careful in how we demarcate the EoR window when calculating spherically-averaged power spectra.
 
\begin{figure*}
	\centering 
	\includegraphics[width=1\textwidth]{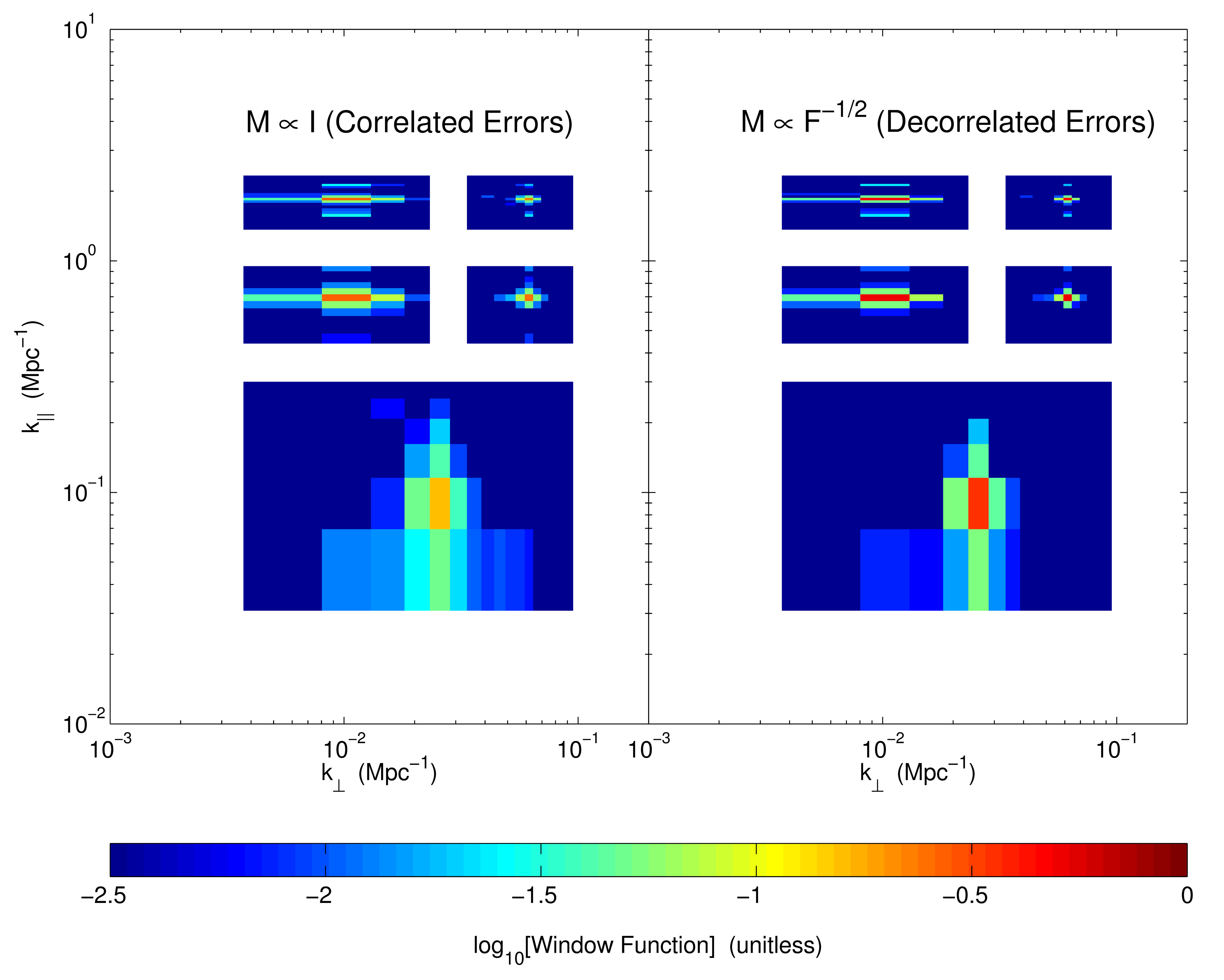}
	\caption{By using an estimator of the 21 cm power spectrum with  uncorrelated errors, we significantly narrow the window functions that relate the ensemble average of our estimator to the true, underlying power spectrum.  Here we show a sample of five cropped window functions for the power spectrum estimate in Figure \ref{fig:2DPkComparison}, each centered at their maxima, for both an estimator with correlated errors (left panel) and an estimator with uncorrelated errors (right panel).  Though the estimator with correlated errors produces smaller vertical error bars, it acheives this by ``over-smoothing" many band powers together.  Narrow window functions let us independently measure many modes of the power spectrum. The band power measured with $\M \sim \F^{-1/2}$ is one of a set of mutually exclusive and collectively exhaustive pieces of information.} 
	\label{fig:2DWindowComparison}
\end{figure*}

In addition to producing a cleaner EoR window, the decorrelated estimator of the power spectrum yields another advantage: narrower window functions.  Both the estimator with the minimum variance and estimator with decorrelated errors represent, in aggregate, the weighted average of the true, underlying band power spectrum, as we discussed in Section \ref{sec:IdealObs}. In Figure \ref{fig:2DWindowComparison}, we show the improvement that the decorrelated estimator offers over the minimum variance estimator by narrowing the window functions considerably.\footnote{While the choice of $\M \sim \F^{-1/2}$ ensures that the power spectrum estimator covariance is diagonal (recall, $\boldsymbol \Sigma = \M \F \M^t$ while $\mathbf{W} = \M \F$), it does not mean that the window functions are delta functions.  The off-diagonal terms of $\boldsymbol \Sigma$ might be zero even if the off-diagonal terms of $\mathbf{W}$ are not.}
We show five example window functions from the same subband that we plot in Figure \ref{fig:2DPkComparison}, cropped to fit together on one set of axes, each centered at their respective peaks.  Because the window functions are normalized to sum to 1, the breadth of each window function is reflected by the value of the central peak.  As we expected, the window functions are considerably narrower for our decorrelated power spectrum estimator.

\begin{figure}[!ht] 
	\centering 
	\includegraphics[width=.48\textwidth]{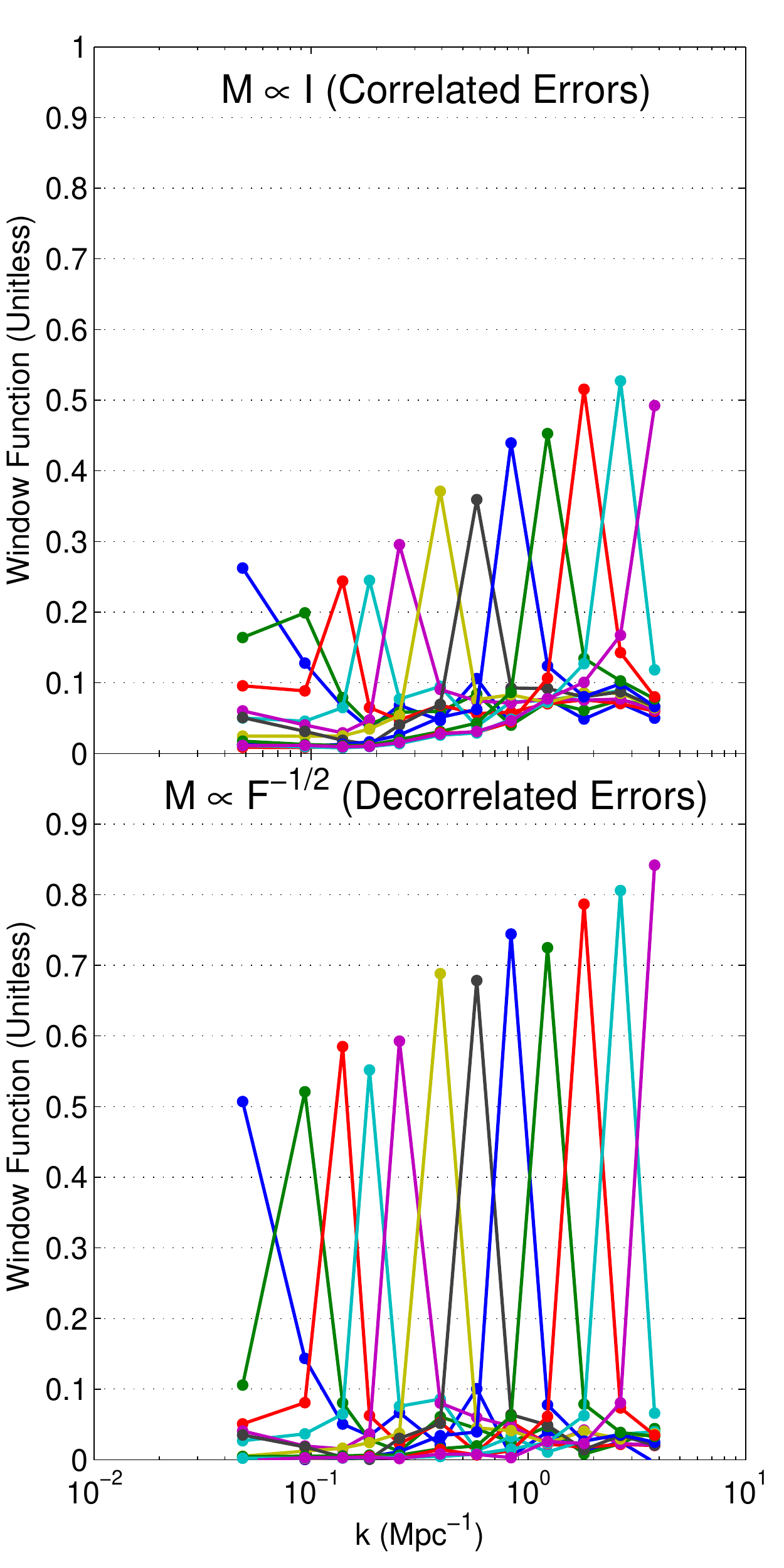}
	\caption{Even after optimally binning the cylindrical power spectra from Figure \ref{fig:2DPkComparison} to spherical power spectra, the choice of a power spectrum estimator with decorrelated errors produces much narrower window functions than the minimum variance technque.  In addition to maintaining a clean EoR window, the choice of $\M \sim \F^{-1/2}$ provides the additional benefit of allowing power spectrum modes to be measured more independently.}
	\label{fig:1DWindowComparison}
\end{figure}

Even after binning from cylindrical power spectra to spherical power spectra, the difference remains quite stark.  In Figure \ref{fig:1DWindowComparison} we see clearly that choosing a power spectrum estimator with decorrelated errors also considerably improves the window functions in one dimension as well as two.

Lastly, as we mentioned in Section \ref{sec:cylindToSph}, one of advantage of our method is that it keeps a full accounting of the error covariance, $\mathbf{\Sigma}$. When $\mathbf{M}$ is not chosen to make $\mathbf{\Sigma}$ diagonal, an improper accounting can lead to a suboptimal or simply incorrect propagation of errors.  In Appendix \ref{sec:ErrorCovAppendix} we work through an example of the consequences of assuming the independence of errors at various steps in the analysis.  This should serve as a warning of the importance of careful analysis; incorrectly assuming a diagonal $\mathbf{\Sigma}$ can lead to unnecessarily wide window functions, an overestimation of errors, or---worst of all---an underestimation of errors that could lead to an unjustified claim of a detection.

 
\section{Early results}
\label{sec:earlyResults}
Having developed and demonstrated a technique that robustly preserves the EoR window while thoroughly and honestly keeping track of the errors on and correlations between our band power estimates, we can now confidently generate some interesting preliminary science results.  Because these data span the widest redshift range to date, we are able to investigate the behavior of the wedge feature over many frequencies.  Understanding the behavior of the EoR window over a large redshift range is important, since there is still considerable uncertainty about the timing and duration of the EoR.  Moreover, it is often argued that a tentative first detection of the cosmological signal will only be convincingly distinguishable from residual foregrounds if one can show that the $21\,\textrm{cm}$ brightness temperature fluctuations peak at some redshift, since theory predicts that the midpoint of reionization should be marked by such a peak \citep{LidzRiseFall,BittnerLoeb}.  It is therefore essential to characterize the EoR window (and by extension, residual foregrounds) over a broad frequency range. We also apply our methods from Section \ref{sec:Methods} to calculate spherically averaged power spectra over our entire redshift range, including error bars and window functions, thus setting a limit on the 21 cm brightness temperature power spectrum during the EoR.

\subsection{The wedge} 
\label{sec:WedgeResults}

\begin{figure*} [!ht] 
	\centering 
	\includegraphics[width=1\textwidth]{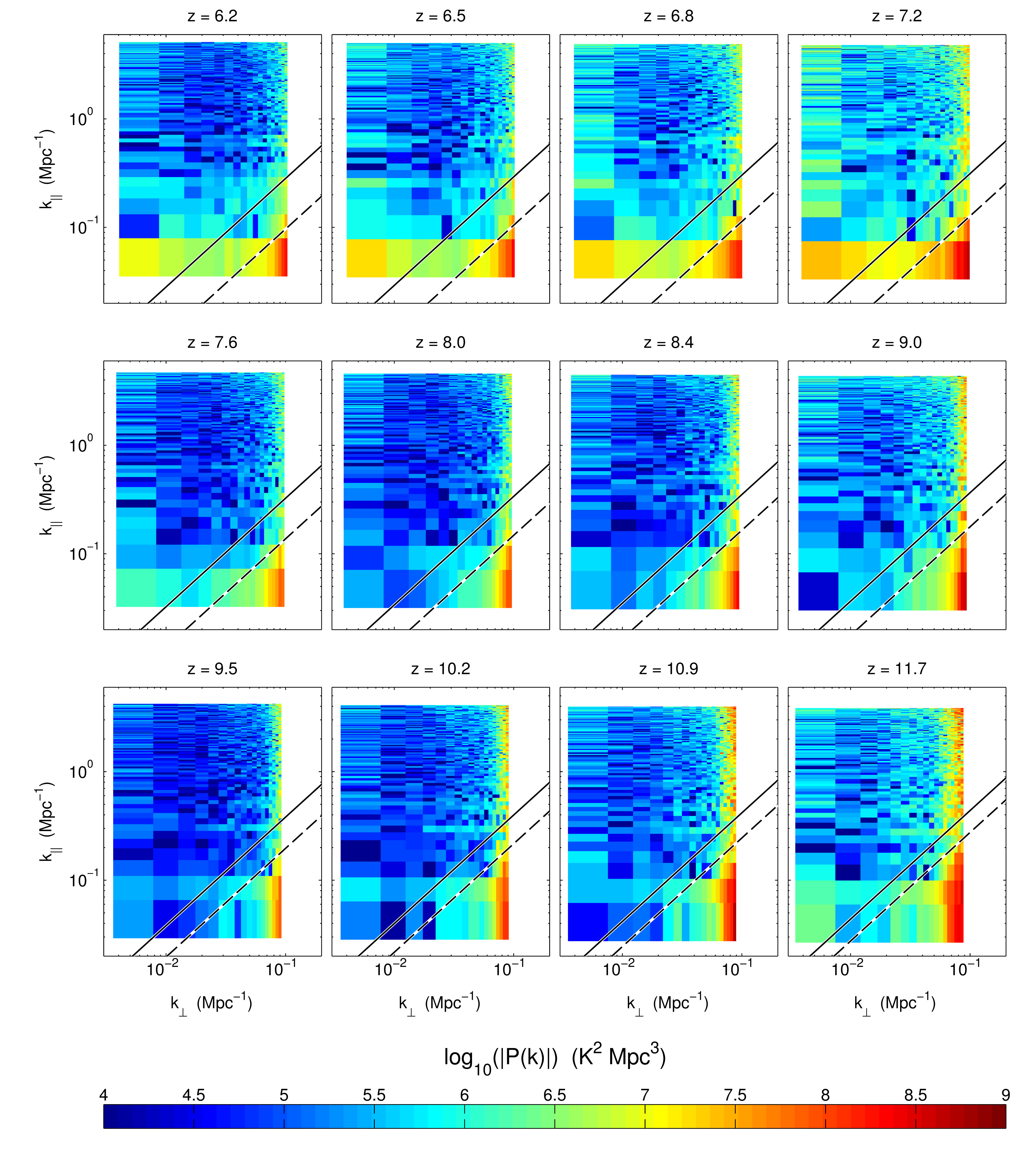}
	\caption{Examining cylindrically binned power spectra for each subband (each averaged over all nine subfields), reveals several important trends with frequency of the EoR window and the foregrounds.  Each row is a single simultaneously observed frequency band.  Since different bands were observed for different amounts of time, direct comparisons between rows is challenging.  However, several clear trends emerge.  For each band, moving to higher redshift (increasing wavelength) shows stronger foregrounds, a larger wedge (in part due to a wider primary beam), and a noisier EoR window (due to a higher system temperature).  In general the brightest foreground contamination is well demarcated by the wedge line in Equation \eqref{eq:wedge} for the primary beam (dotted line) and especially by the wedge line for the horizon (solid line).  In short, the wedge displays the theoretically expected frequency dependence.}
	\label{fig:allWedges}
\end{figure*}

In Figure \ref{fig:allWedges}, we show all the cylindrical power spectra over the redshift range probed by our current observations.  
The spectra are sorted into three rows, each of which contain data coming from a single $30.72\,\textrm{MHz}$ wide frequency band.  All of the spectra were generated using the same techniques that were used to generate the example cylindrical power spectra in Section \ref{sec:MethodDemo} and thus contain all the desirable statistical properties discussed in Section \ref{sec:Methods}.  One sees that in every case the foregrounds are mostly confined to the wedge region in the bottom right corner of the $k_\perp$-$k_\parallel$ plane.  This builds upon the single frequency observations of \cite{PoberWedge}, demonstrating the existence of the EoR window across a wide range of frequencies relevant to EoR observations.

Having these measurements also allows us to examine the behavior of the EoR window as a function of frequency.  Consider first the high $k_\perp$ regions of the $k_\perp$-$k_\parallel$ plane.  The most striking feature here is the wedge.  Consistent with being dominated by foreground power, the wedge generally gets brighter with decreasing frequency within each wide frequency band, just as foreground emission is known to behave.  The extent of the wedge is also in line with theoretical expectations.  Recall from Equation \eqref{eq:wedge} that the wider the field-of-view, the farther up in $k_\parallel$ the wedge goes.  Since the field-of-view is defined by the primary beam, whose extent decreases with increasing frequency, one expects the wedge to have the largest area at the lowest frequencies.  This trend is clearly visible in the cylindrical power spectra of Figure \ref{fig:allWedges}, where the wedge extends to the highest $k_\parallel$ at the highest redshifts.  Importantly, the wedge is confined to its expected location across the entire range of the observations.  To see this, note that we have overlaid Equation \eqref{eq:wedge} on the plots, with the dashed line corresponding to $\theta_{\textrm{max}}$ equal to that of the first null of the primary beam, and the solid line with $\theta_{\textrm{max}} = \pi / 2$ (the horizon).  At all frequencies, the most serious contaminations lie within the first null, ensuring that the EoR window is foreground-free.

Foregrounds also enter indirectly into the instrumental noise-dominated regions because the MWA is sky-noise dominated.  Thus, as the brightest sources of emission in our observations, the foregrounds set the system temperature, and result in a higher instrumental noise at higher redshifts.  This trend can be seen within each wide frequency band (each row of Figure \ref{fig:allWedges}), although the slight interruption of this trend between bands suggests an additional source of noise.

At low $k_\perp$, theory suggests that foregrounds will contaminate a horizontally-oriented region at the bottom of the plot.  This is clearly seen in the highest frequency plots.  Interestingly, at lower frequencies the increasing instrumental noise plays more of a role, and the foreground contribution is less obvious in comparison (although it is still there).  While a naive reading of some of these low frequency plots (such as the one for $z=9.1$) might suggest that the EoR window extends to the lowest $k_\parallel$, such a conclusion would be misguided.  As we shall see in Section \ref{sec:Limits}, these modes are likely dominated by foregrounds (and therefore do not integrate down with further integration unlike instrumental noise dominated modes).  Moreover,  the error statistics (which self-consistently include foreground errors in our formalism) suggest that low $k_\parallel$ modes are less useful for constraining theoretical models, and that the true EoR window does in fact lie at higher $k_\parallel$, as suggested by theory.  Again, this highlights the importance of estimating power spectra in a framework that naturally contains a rigorous calculation of the errors involved.

\subsection{Spherical Power Spectrum Limits}
\label{sec:Limits}

\begin{figure*} [!ht] 
	\centering 
	\includegraphics[width=1\textwidth]{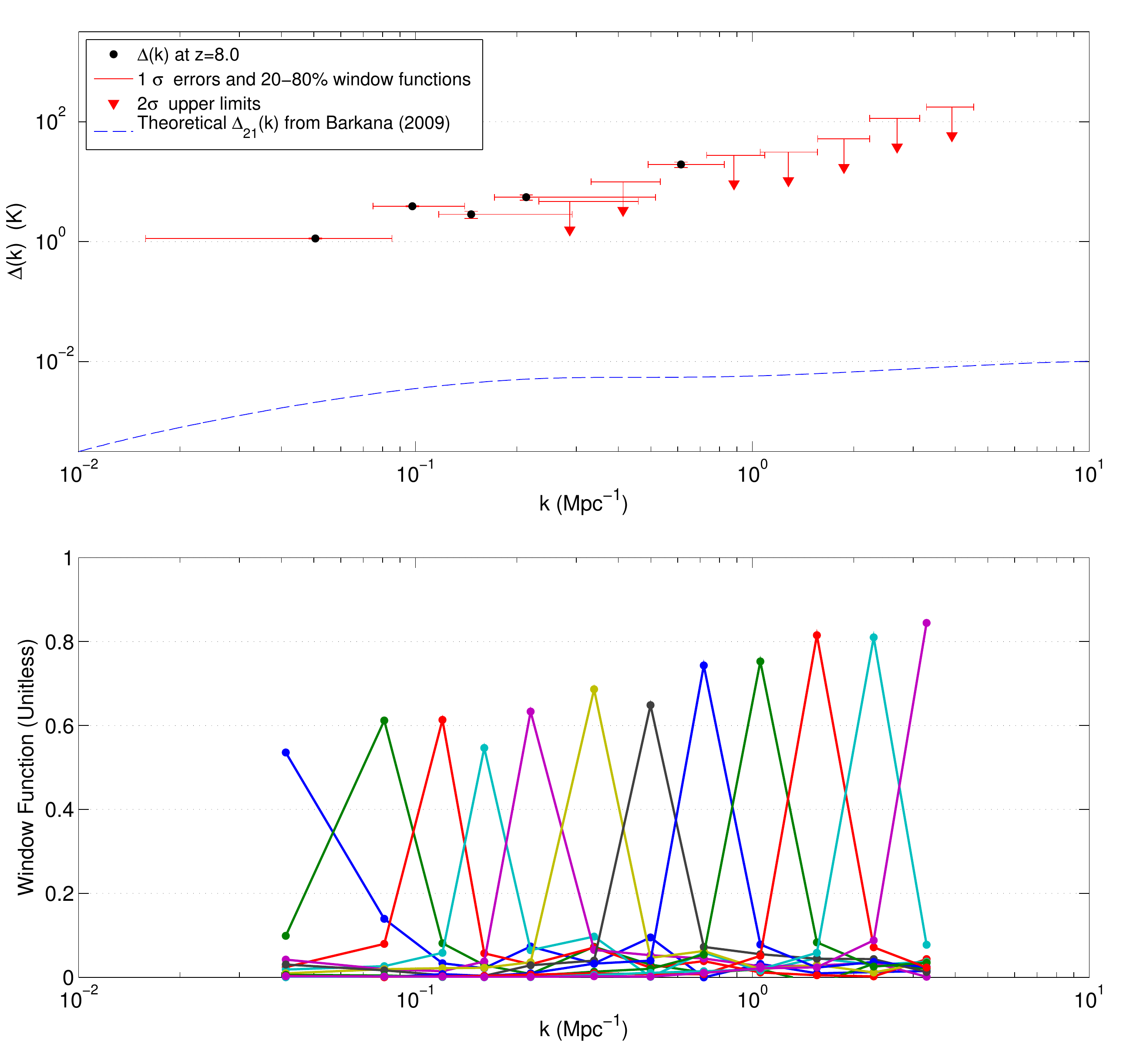}
	\caption{Our method allows for the estimation of the spherically binned power spectrum in temperature units, $\Delta(k)$, while keeping full acount of both vertical error bars and window functions (horizontal error bars) and making an optimal choice in the tradeoff between the two.  In the top panel, we have plotted our spherical power spectrum estimates of the subband centered on 158 MHz ($z = 8.0$), including $1\sigma$ errors on detections (which are often only barely visible), $2\sigma$ upper limits on non-detections, and horizontal error bars that span the middle three quintiles of the window functions (bottom panel).  At low $k$, the wide error bars are the expected consequence of foreground contamination \citep{LT11}.  Downward arrows represent measurements consistent with noise at the $2\sigma$ level. Even though the area under the primary beam wedge has been excised from the 2D-to-1D binning, the detection of foregrounds at low $k$, is expected due to the contribution of unresolved foregrounds over a wide range of $k_\perp$ \citep{DillonFast}. Our fiducial theoretical power spectrum is taken from \cite{BarkanaPS2009}.}
	\label{fig:singleDelta}
\end{figure*}

Having confirmed that the EoR window behaves as expected, we will now proceed to place constraints on the spherical power spectrum.  In top panel of Figure \ref{fig:singleDelta} we show the result of binning the $z=10.3$ cylindrical power spectrum of Figure \ref{fig:allWedges}, using the optimal binning formulae presented in Section \ref{sec:cylindToSph}.
In addition, for ease of interpretation, we elect to plot
\begin{equation}
\Delta (k) \equiv \sqrt{ \frac{k^3}{2 \pi^2} P(k)}
\end{equation}
(which simply has units of temperature) rather than $P(k)$ itself.  

To quantify the errors in our spherical power spectrum estimate, we also bin the cylindrical power spectrum measurement covariances and window functions using the formulae of Section \ref{sec:cylindToSph}.  The resulting window functions are shown in the bottom panel, and give an estimate of the horizontal error bars.  Thinking of these window functions (which, recall, are normalized to integrate to unity) as probability distributions, the horizontal error bars shown in the top panel are demarcated by the 20th and 80th percentiles of the distribution.  (This corresponds to the full-width-half-maximum in the event that the window functions are Gaussians).  The vertical error bars were obtained by taking the square root of each diagonal element of the covariance matrix.  Since the methods of Section \ref{sec:cylindToSph} carefully preserved the diagonal nature of the bandpower covariance, each data point in Figure \ref{fig:singleDelta} represents a statistically independent measurement.  This would not have been the case had we not employed the decorrelation technique of Section \ref{sec:decorr}.

Immediately obvious from the plot is that there is a qualitative difference between the data points at low $k$ and those at high $k$.  In particular, the points at low $k$ are detections of the sky power spectrum, whereas the points at high $k$ are formal upper limits. This is not to say, of course, that the cosmological EoR signal has been detected at low $k$.  Rather, recall from Section \ref{sec:crossPower} that in an attempt to avoid having to make large bias subtractions, we elected to compute cross-power spectra of total sky emission rather than of the cosmological signal, with the expectation (largely confirmed in Section \ref{sec:WedgeResults}) that the intrinsic cleanliness of the EoR window would be sufficient to ensure a relatively foreground-free measurement at high $k_\parallel$.  Now, our survey volume is such that we are sensitive almost exclusively to regions in Fourier space where $k_\parallel \gg k_\perp$.  When binning along contours of constant $k$ in the cylindrical Fourier space, we have that $k \equiv \sqrt{k_\perp^2 + k_\parallel^2} \approx k_\parallel$, and therefore the low $k$ points of Figure \ref{fig:singleDelta} map to low $k_\parallel$.  The detections seen at low $k$ thus reside outside the EoR window and are almost certainly detections of the foreground power spectrum.  

Despite the fact that the low $k$ modes are foreground dominated, they still constitute a formal upper limit on the cosmological power spectrum, since the foreground power spectrum is necessarily positive.  In fact, our current, most competitive upper limit resides at the lowest $k$ values.  However, this is unlikely to continue to be the case as more data is taken with the MWA, for two reasons.  First, as foreground-limited measurements, the data points at low $k$ will not average down with further integration time.  In addition, the error statistics in the region are not particularly encouraging.  The window functions (and therefore the horizontal error bars) are seen to broaden towards lower $k$, reducing the ability of constraints at those $k$ to place limits on theoretical models.  (This is most easily seen by recalling that the window functions integrate to unity by construction, and thus the increase in their peak values towards higher $k$ implies a broadening of the window functions).  The broadening of the window functions is an expected consequence of foreground subtraction \citep{LT11} and thus will likely continue to limit the usefulness of the low $k$ regime unless future measurements can characterize foreground properties with exquisite precision.

In contrast, the points at high $k$ do reside in the EoR window.  The constraints here are limited by thermal noise, as we saw in Section \ref{sec:MethodDemo}.  Bolstering this view is the fact that the data here are consistent with zero, as one expects for a noise-dominated cross-power spectrum.  The limits here are given by the $2\sigma$ errors predicted by the Equation \eqref{eq:Cylind2SphCovar}.  As mentioned in Section \ref{sec:CovarModeling}, these errors are somewhat larger than what might be predicted by a theoretical sensitivity calculation.  However, they are consistent with rough estimates of the errors obtained from a calculation of root-mean-square values from the images produced in Section \ref{sec:Mapmaking}.  This suggests that the larger-than-expected errors are due to noisier-than-expected input maps, and not to any approximations made in the power spectrum estimation techniques presented in this paper.  The data on which these results are based are from the very first operation of the prototype array, and we expect better performance in later data.  Encouragingly, we note also that as noise-dominated constraints, the measurements at high $k$ will continue to improve with integration time. 
 
\begin{figure*} 
	\centering 
	\includegraphics[width=1\textwidth]{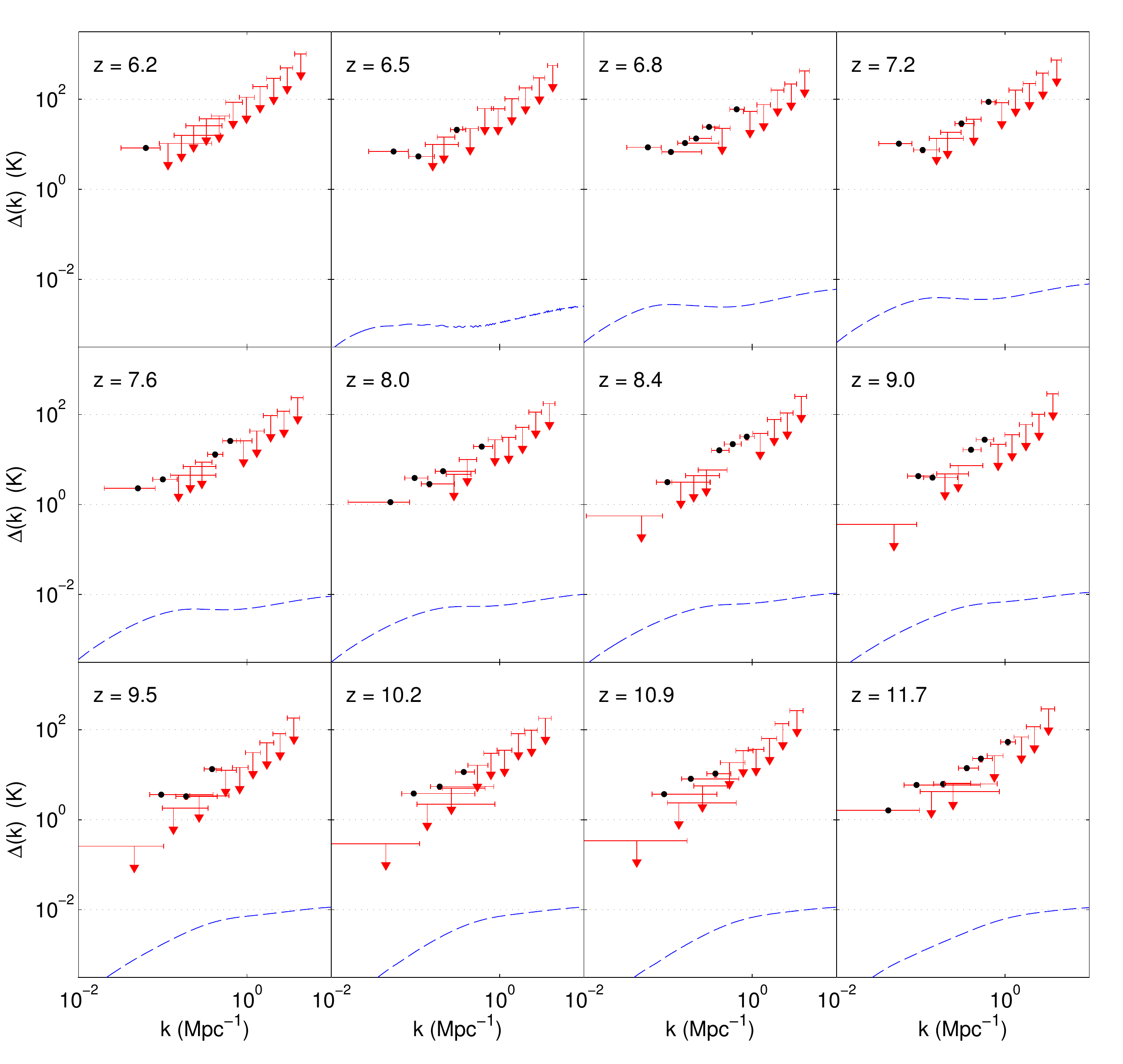}
	\caption{Taking advantage of our fast yet thorough power spectrum estimation technique, we estimate $\Delta(k)$ for a wide range of $k$ and $z$, including both vertical and horizontal errors.  (For points that represent positive detections of foregrounds or systematic correlations, the vertical error bars are often barely visible).  Using the visual language of Figure \ref{fig:singleDelta}, we show here our spherical power spectrum limits as a function of both $k$ and $z$.  Each panel is a different subband.  The many detections can be attibuted to foregrounds (especially at low $k$), instrumental effects like those we saw in Figure \ref{fig:2DPkComparison} (especially at medium values of $k$), or both.	Our absolute lowest limit on the 21 cm brightness temperature power spectrum, $\Delta(k) < 0.3$ Kelvin at the 95\% confidence level, comes at $k=0.046$ cMpc$^{-1}$ and $z=9.5$ (or $\Delta (k) < 2 \,\textrm{K}$ at $z=9.5$ and $k= 0.134\,\textrm{cMpc}^{-1}$ if one discards the lowest $k$ bin to immunize oneself against foreground modeling uncertainties).}
	\label{fig:allDelta}
\end{figure*}

In Figure \ref{fig:allDelta}, we show power spectrum limits across the entire frequency range of the MWA, along with some theoretical predictions generated using the models in \cite{BarkanaPS2009}.  At the lowest redshift, no theory curve is plotted because the model predicts that reionization is complete by then.  This yet again underscores the importance of making measurements over a broad frequency range---with access to a wide range of redshifts, future detections of the cosmological signal can be distinguished from residual foregrounds by measuring null signals at redshifts where reionization is complete.

Each redshift bin of Figure \ref{fig:allDelta} exhibits trends that are qualitatively similar to those discussed above for the $z=10.3$ case.  We see many apparent detections of correlations positive correlations between the two time-interleaved data cubes---more than can be attributed to foregrounds alone.  As we saw with the cylindrically binned power spectrum in Figure \ref{fig:2DPkComparison}, there is evidence of systematic and instrumental effects sending foreground power into the EoR window, leading to higher $k$ detections and large differences between neighboring $k$ bins.  With as new an instrument as the MWA was at time of this observation, this issues are understandable.  The exact physical origin of those systematics is beyond the scope of this paper, however they should serve as a reminder to stay vigilant for them in future datasets from a more battle-tested instrument.  However, because we see no evidence of strong anti-correlations between data cubes, we expect that the extra power introduced by systematics into the EoR window only the effect of worsening the limits we can set.  

Over all bands, our best limit is $\Delta (k) < 0.3 \,\textrm{K}$, occurring at $z=9.5$ and $k= 0.046\,\textrm{cMpc}^{-1}$.  However, as remarked in Section \ref{sec:CovarModeling}, the lowest $k$ bins can be rather sensitive to the covariance model, and if one excludes those bins, our best limit is $\Delta (k) < 2 \,\textrm{K}$, at $z=9.5$ and $k= 0.134\,\textrm{cMpc}^{-1}$.  While our limits may not be quite as low as other existing limits in the literature \citep{newGMRT,AaronInPrep}, they are the only limits on the EoR power spectrum that span a broad redshift range from $z=6.2$ to $z=11.7$.  Moreover, these statistically rigorous limits will likely improve with newer and more sensitive data from the MWA.

\section{Conclusions}
\label{sec:Conc}
In this paper, we have accomplished three goals.  First, we adapted $21\,\textrm{cm}$ power spectrum estimation techniques from \citet{LT11} and \citet{DillonFast} with real-world obstacles in mind, so that they could be applied to real data.  With early MWA data, our generalized formalism was then used to demonstrate the importance of employing a statistically rigorous framework for power spectrum estimation, lest one corrupt the naturally foreground-free region of Fourier space known as the EoR window.  Finally, we used the MWA data to set limits on the EoR power spectrum.

In confronting real-world obstacles, our desire is to preserve the as much of the statistical rigor in previous matrix-based power spectrum estimation frameworks as possible.  To avoid having to perform direct subtractions of instrumental noise biases, we advocate computing cross-power spectra between statistically identical subsets of the data (in the case of the MWA worked example of this paper, these subsets were formed from odd and even time samples of the data).  This has the effect of eliminating noise bias in the power spectrum, although instrumental noise continues to contribute to the error bars.  To avoid direct subtractions of foreground biases, we simply look preferentially in the EoR window, where foregrounds are expected to be low.  Missing data, whether from incomplete $uv$ coverage or RFI flagging, can be dealt with using the pseudoinverse formalism.  Doing this allows the effects of missing data to be self-consistently propagated into error statistics such as the power spectrum covariance and the window functions.  In an effort to preserve the cleanliness of the EoR window, one should form decorrelated bandpower estimates, which have uncorrelated errors and reasonably narrow window functions.  Care must then be taken to preserve these nice properties via an optimal binning of cylindrical bandpowers into spherical bandpowers.

Using early MWA data to demonstrate these techniques, we have confirmed theoretical predictions for the existence of the EoR window and have extended previous observations done by other groups to a much wider frequency range.  This allowed us to check predicted trends of the EoR window as a function of frequency, all of which are consistent with theory.  Crucially, we found that without using the decorrelation technology of Section \ref{sec:decorr}, measurements in the EoR window are not in fact instrumental noise dominated, and contain a systematic bias that is indicative of foreground leakage from outside the EoR window.

The early MWA data has also allowed us to place limits on the cosmological EoR power spectrum.  Our best limit is $\Delta (k) < 0.3\,\textrm{K}$, at $z=9.5$ and $k=0.046\,\textrm{cMpc}$ (or $\Delta (k) < 2 \,\textrm{K}$ at $z=9.5$ and $k= 0.134\,\textrm{cMpc}^{-1}$ if one discards the lowest $k$ bin to immunize oneself against foreground modeling uncertainties).  This may not be competitive with other published observations, but generalizes them in an important way: instead of focusing on one particular frequency, our limits span a wide range of redshifts relevant to the EoR, going from $z=6.2$ to $z=11.7$.  In addition, these limits will almost certainly improve in the near future, using already-collected (but yet to be analyzed) data from the MWA-32T system, as well as soon-to-be-collected data from the MWA-128T system.  The rigorous statistical tools developed in this paper should be equally applicable to these newer data sets, ensuring that foreground contamination remains confined to outside the EoR window, safeguarding the potential of current generation experiments to make an exciting first detection of the EoR within the next few years.

\section*{Acknowledgments}
This work makes use of the Murchison Radio-astronomy Observatory. We acknowledge the Wajarri Yamatji people as the traditional owners of the Observatory site. Support for the MWA comes from the U.S. National Science Foundation (grants AST-0457585, PHY-0835713, CAREER-0847753, and AST-0908884), the Australian Research Council (LIEF grants LE0775621 and LE0882938), the U.S. Air Force Office of Scientic Research (grant FA9550-0510247), and the Centre for All-sky Astrophysics (an Australian Research Council Centre of Excellence funded by grant CE110001020). Support is also provided by the Smithsonian Astrophysical Observatory, the MIT School of Science, the Raman Research Institute, the Australian National University, and the Victoria University of Wellington (via grant MED-E1799 from the New Zealand Ministry of Economic Development and an IBM Shared University Research Grant). The Australian Federal government provides additional support via the National Collaborative Research Infrastructure Strategy, Education Investment Fund, and the Australia India Strategic Research Fund, and Astronomy Australia Limited, under contract to Curtin University. We acknowledge the iVEC Petabyte Data Store, the Initiative in Innovative Computing and the CUDA Center for Excellence sponsored by NVIDIA at Harvard University, and the International Centre for Radio Astronomy Research (ICRAR), a Joint Venture of Curtin University and The University of Western Australia, funded by the Western Australian State government.  Additionally, the authors wish to thank Aaron Ewall-Wice, Lu Feng, Abraham Neben, Aaron Parsons, Jonathan Pober, Ronald Remillard, and Richard Shaw for valuable discussions, and Rennan Barkana for both useful discussions and for providing the theory results shown in Figures \ref{fig:singleDelta} and \ref{fig:allDelta}.  This work is partially supported by NSF grants AST-0821321 and AST-1105835.  A.L. acknowledges support from the Berkeley Center for Cosmological Physics.

\appendix

\section{On the Importance of Modeling the Full Error Covariance}
\label{sec:ErrorCovAppendix}

\begin{figure*}[!ht] 
	\centering 
	\includegraphics[width=1\textwidth]{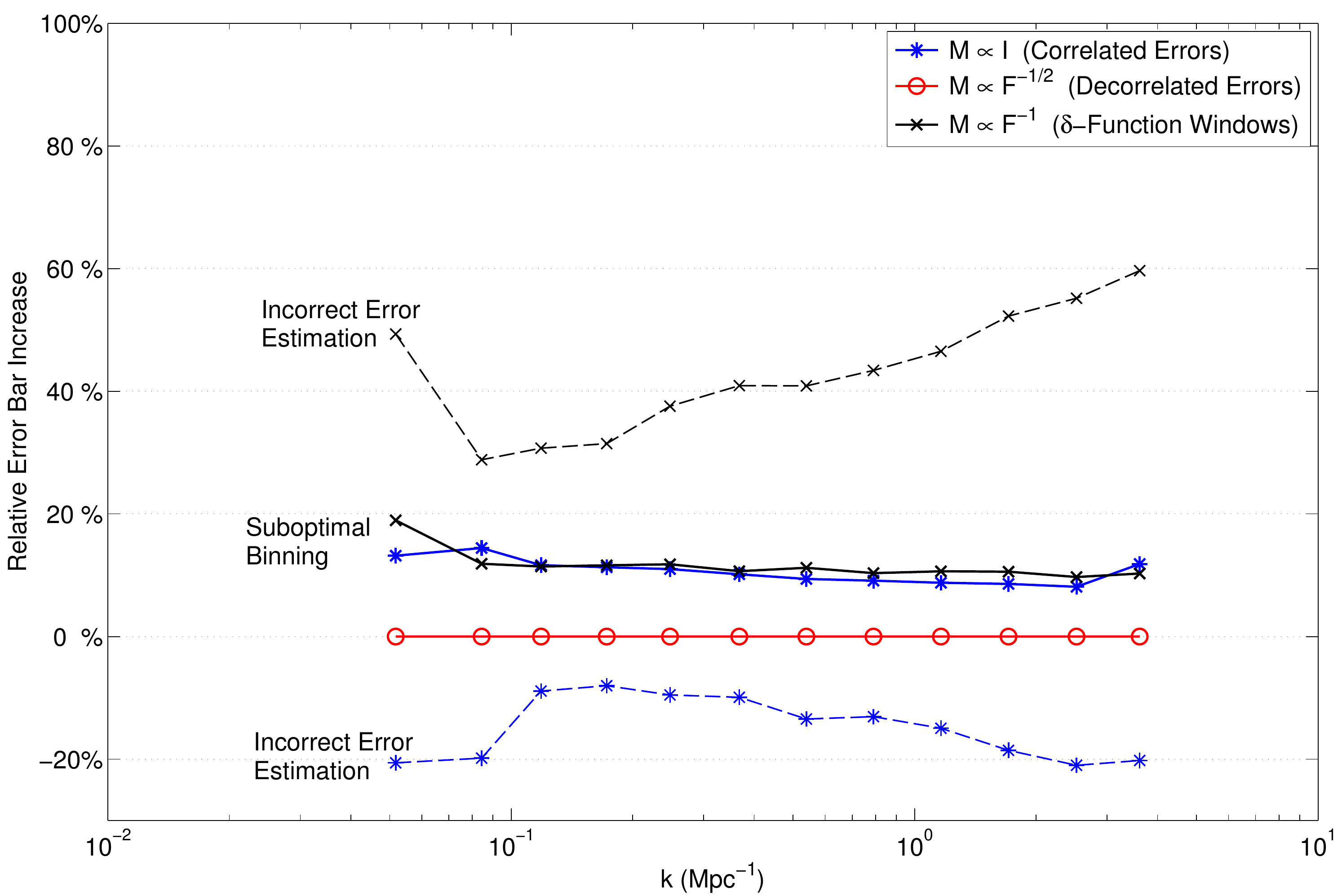}
	\caption{Neglecting the fact that the covariance of the power spectrum estimator is, in general, non-diagonal, can lead to two mistakes that can either unnecessarily enlarge our error bars or, even worse, unjustifiably shrink them.  In this figure, we first show an approximately 10\% increase in the vertical error bars on the power spectrum (solid lines) from a suboptimal inverse variance weighted binning scheme, rather than the inverse covariance weighted binning of Equation \eqref{eq:Cylind2SphEst}.  This problem is obviated by choosing an estimator with decorrelated errors and thus a diagonal covariance matrix.  If one simply assumes that the estimator covariance in Equation \eqref{eq:Cylind2SphCovar} is diagonal when it is not (dotted lines), one is led, depending on the choice of estimator, either to roughly 50\% larger error bars than necessary or, worse yet, artificially small error bars.  The last mistake, choosing an estimator with small error bars---despite its wide window functions---and then neglecting the off-diagonal terms in the estimator covariance, is potentially the most pernicious since it could lead to a claimed detection in the absence of signal.}
	\label{fig:VarVsCovarErrors}
\end{figure*}

In Section \ref{sec:cylindToSph}, we argued that an inverse covariance weighted binning scheme for estimating spherical band powers produced optimal spherical power spectrum estimate.  In the case where $\M$ is chosen either for the smallest possible error bars or the narrowest possible window functions, the estimator covariance $\boldsymbol \Sigma^\text{cyl}$ is non-diagonal.  Assuming that the matrix is actually diagonal, at one or more steps in the binning and error propagation, can lead error bars that are overly conservative or---worse yet---error bars that are insufficiently conservative and might falsely lead to a claimed detection.  In Figure \ref{fig:VarVsCovarErrors}, we show the effects of making a suboptimal choice for binning.

If one fully models the covariance matrix $\boldsymbol \Sigma^\text{cyl}$, including off-diagonal elements, but chooses to generate $\widehat{\mathbf{p}}^\text{sph}$ as an inverse variance (and not inverse covariance) weighted average of cylindrical band powers, neglecting off diagonal terms in the weighting, one's estimators will be noisier as a result (see the solid lines in Figure \ref{fig:VarVsCovarErrors}). These are the correct errors for the suboptimal choice of estimators. 

\begin{figure} [!ht] 
	\centering 
	\includegraphics[width=.5\textwidth]{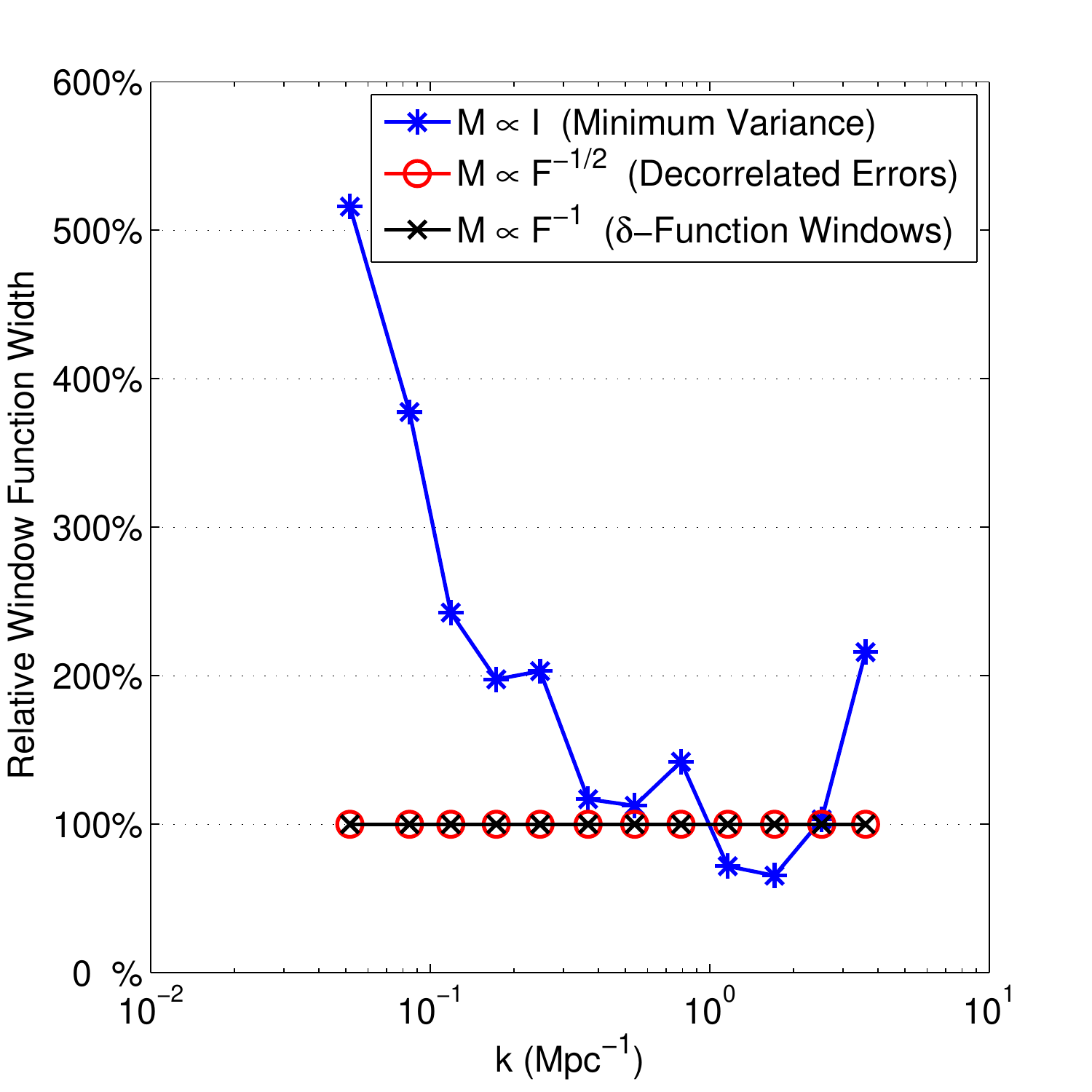}
	\caption{Just as with the error bars in Figure \ref{fig:VarVsCovarErrors}, generating suboptimally binned spherical power spectrum estimates by neglecting off-diagonal terms in the estimator covariance can lead to wider window functions than necessary.  We illustrate the effect by comparing the width of the window functions between the 20th and 80th percentiles between the two binning schemes. This is important for the choice of power spectrum estimator with the smallest error bars and widest window functions ($\M \sim \Eye$).  In the case where our power spectrum estimator has uncorrelated errors, there are no off-diagonal terms in the estimator covariance and both binning schemes are identical.  In the case of the estimator with $\delta$-function window functions, suboptimal binning does not affect the window functions---though it still affects the vertical errors (see Figure \ref{fig:VarVsCovarErrors}).}
	\label{fig:VarVsCovarWindows}
\end{figure}

Even worse, if one assumes that $\boldsymbol \Sigma^\text{cyl}$ is diagonal when it is not, one is led either to overestimate the error bars, in the case of $\M \sim \F^{-1}$, or underestimate them, as would be the case when $\M \sim \Eye$.  This is because the former case general exhibits anti-correlated errors while the latter suffers from correlated errors.  The last scenario is the most troubling: by aggressively choosing the estimator with the smallest vertical error bars ($\M \sim \Eye$) and then neglecting the correlations between errors, one will underestimate the error bars and might be lead to falsely claiming a detection.  In this case, the estimator is suboptimal and the errors are incorrect.  Additionally, as we show in Figure \ref{fig:VarVsCovarWindows}, if one were to calculate the the window functions under the assumption that $\boldsymbol \Sigma^\text{cyl}$ is diagonal, one would find window functions several times boarder than they would otherwise be.  
  
Thankfully, choosing the cylindrical power spectrum estimator with decorrelated errors avoids the subtle difference between inverse variance and inverse covariance weighted binning.  The $\M \sim \F^{-1/2}$ decorrelated estimator preserves the EoR window and allows for easy, optimal binning of uncontaminated regions into spherical band power spectrum estimates.

\bibliography{PSPaper}  

\end{document}